\documentclass[]{aa}
\usepackage{linenoaa}
\providecommand{\dodoi}[1]{doi:~\href{http://doi.org/#1}{\nolinkurl{#1}}}
\providecommand{\doarXiv}[1]{arXiv:~\href{https://arxiv.org/abs/#1}{\nolinkurl{#1}}}
\nolinenumbers
\DeclareRobustCommand{\VAN}[3]{#2}
\let\VANthebibliography\thebibliography
\def\thebibliography{\DeclareRobustCommand{\VAN}[3]{##3}\VANthebibliography}

\usepackage{academicons}
\usepackage{xcolor}
\usepackage{orcidlink}
\usepackage{tablefootnote}
\usepackage{subfig}

\def\arcsec{\hbox{$^{\prime\prime}$}}
\def\approxlt{\ifmmode \rlap{$<$}{}_{{}_{{}_{\textstyle\sim}}} \else%
$\rlap{$<$}{}_{{}_{{}_{\textstyle\sim}}}$\fi}

\def\xmm{{\it XMM-Newton}}
\def\swf{{\it Swift}}

\def\nus{{\it NuSTAR}}

\def\arcsec{\hbox{$^{\prime\prime}$}}

\graphicspath{{./}{figures/}}

\begin{document}
\titlerunning{Slim disk modeling reveals an IMBH in the LFBOT AT2018cow}
\authorrunning{Cao et al.}

\title{Slim disk modeling reveals an accreting intermediate-mass black hole in the luminous fast blue optical transient AT2018cow}

\author{Zheng Cao\orcidlink{0000-0002-0588-6555}
\inst{1,2}\fnmsep\thanks{z.cao@sron.nl},
Peter G.~Jonker\orcidlink{0000-0001-5679-0695}\inst{2,1},
Sixiang Wen\orcidlink{0000-0002-0934-2686}\inst{3},
\and
Ann I.~Zabludoff\orcidlink{0000-0001-6047-8469}\inst{4}}

\institute{SRON, Netherlands Institute for Space Research, Niels Bohrweg 4, 2333 CA, Leiden, The Netherlands
\and
Department of Astrophysics/IMAPP, Radboud University, P.O.~Box 9010, 6500 GL, Nijmegen, The Netherlands
\and
National Astronomical Observatories, Chinese Academy of Sciences, 20A Datun Road, Beijing 100101, China
\and
University of Arizona, 933 N. Cherry Ave., Tucson, AZ  85721
}






 \date{Received XXX; accepted XXX}
\abstract{
The origin of the most luminous subclass of the fast blue optical transients (LFBOTs) is still unknown. We present an X--ray spectral analysis of AT2018cow---the LFBOT archetype---using \textit{NuSTAR}, \textit{Swift}, and \textit{XMM-Newton} data. The source spectrum can be explained by the presence of a slim accretion disk, and we find that the mass accretion rate decreases to sub--Eddington levels $\gtrsim$ 200~days after the source's discovery. Applying our slim disk model to data obtained at multiple observational epochs, we constrain the mass of the central compact object in AT2018cow to be log($M_{\bullet}/M_{\odot}$)$=2.4^{+0.6}_{-0.1}$ at the 68\% confidence level. Our mass measurement is independent from, but consistent with, the results from previously employed methods. The mass constraint is consistent with both the tidal disruption and the black hole--star merger scenarios, if the latter model can be extrapolated to the measured black hole mass. Our work provides evidence for an accreting intermediate--mass black hole ($10^{2}$---$10^{6}~M_{\odot}$) as the central engine in AT2018cow, and, by extension, in LFBOT sources similar to AT2018cow.} 

\maketitle


\nolinenumbers

\section{Introduction}

Fast Blue X--ray Transients (FBOTs; e.g., \citealt{drout2014rapidly,arcavi2016rapidly,tanaka2016rapidly,pursiainen2018rapidly,tampo2020rapidly,ho2022cosmological,ho2023search}) have attracted significant attention in recent years as their physical nature is not yet known. Those at the most luminous end ($\gtrsim10^{44}$~erg~s$^{-1}$) of the FBOT population are often referred to as Luminous FBOTs (LFBOTs). 
LFBOTs are characterized by a rapid optical rise and high peak luminosity (reaching the peak luminosity on a timescale of days; e.g., \citealt{drout2014rapidly,pursiainen2018rapidly,rest2018fast,ho2023search}). 

The archetype of LFBOTs is AT2018cow. AT2018cow was discovered on 2018-06-16 (Modified Julian Date, or MJD, 58285) by the ATLAS survey \citep{smartt2018atlas18qqn}. The host galaxy CGCG137-068 has a luminosity distance of $\sim$60 Mpc (redshift $z=0.01404$; \citealt{adelman2008sixth}). AT2018cow is the nearest LFBOT to Earth and has been observed across a broad energy range. Multi--wavelength observations show that the source emission extends from radio to $\gamma$--rays \citep[e.g.,][]{prentice2018cow,rivera2018x,kuin2019swift,margutti2019embedded,perley2019fast,ho2019at2018cow,nayana2021ugmrt}. In particular, X--ray emission was observed immediately after the source's discovery \citep{rivera2018x,kuin2019swift}. Analysis of the X--ray spectrum shows that the earliest deep X--ray observation of \nus{} could be well--described by reflection off an accretion disk \citep{margutti2019embedded}. 

Different models have been proposed to explain the multi--wavelength behavior of AT2018cow, or AT2018cow--like LFBOTs (e.g., AT2020xnd/ZTF20acigmel, \citealt{perley2021real}; AT2020rmf, \citealt{yao2022x}). One class of scenarios involves an accreting compact object, either a neutron star (NS) or a black hole (BH), as the "central engine" for the highly--variable, non--thermal X--ray emission. Examples in this class of models include i) a tidal disruption event (TDE) involving an intermediate--mass black hole (IMBH, BH mass $M_{\bullet}$ between $10^{2}$ and $10^{6}~M_{\odot}$; \citealt{kuin2019swift,perley2019fast}); ii) a core--collapse event such as a supernova giving birth to the central compact object, which then accretes fall-back progenitor material \citep[e.g.,][]{prentice2018cow,margutti2019embedded,perley2019fast,mohan2020nearby,gottlieb2022shocked}; iii) a binary merger of a BH and its massive stellar companion \citep{metzger2022luminous}. 

Meanwhile, there is another class of models that relies exclusively on shock interactions in the circumstellar material (CSM; e.g., \citealt{rivera2018x,fox2019signatures,leung2021fast,pellegrino2022circumstellar}). However, it is challenging for these models to explain the observed early--time X--ray/$\gamma$--ray behavior without also invoking an accreting compact object (see, e.g., \citealt{margutti2019embedded,coppejans2020mildly,pasham2022evidence,yao2022x,metzger2022luminous,migliori2023roaring}). Furthermore, late--time ($\gtrsim$ 200~days since MJD~59295) observations of AT2018cow reveal a soft X--ray source spectrum \citep{migliori2023roaring}. Moreover, the source enters a long--lasting "plateau" phase in the UV lightcurves \citep[e.g.,][]{inkenhaag2023late}, resembling a BH 
evolving from a high to a low 
mass accretion rate, which
can last months to years (e.g., similar to what has been seen in many TDEs; \citealt{van2019late,mummery2020spectral,wen2023optical,cao2023rapidly,mummery2024fundamental}).

The mass of the central compact object is of key importance to unveil the nature of AT2018cow--like LFBOTs. To that end, it is essential to compare different mass measurements to verify the different measurement methods. Several studies have reported (limits on) the compact object mass in AT2018cow. From an X--ray timing analysis, \citet{pasham2022evidence} find an upper limit on the central object mass of $\lesssim850$~$M_{\odot}$, assuming the quasi--periodic oscillations in the arrival times of X--ray photons are due to particular orbital frequencies in an accretion disk (but see \citealt{zhang2022possible}). \citet{migliori2023roaring} constrain the compact object mass to be $\approx$~10~--~10$^{4}$~$M_{\odot}$ based on energetic arguments, while \citet{inkenhaag2023late} find the mass to be 10$^{3.2\pm0.8}$~$M_{\odot}$ based on modeling of the late--time UV emission as coming from a TDE--like accretion disk. In this paper, we use X--ray spectral analysis to provide a mass measurement of the central compact object.

The paper is structured as follows: In Section 2, we describe the data and the data reduction method. In Section 3, we describe the slim disk model. In Section 4, we present the results from our analysis. In Section 5 we discuss the results and present our conclusions.

\section{Methods and data reduction}

In this study, we use Poisson statistics (\citealt{cash1979parameter}; {\sc C-STAT} in {\sc XSPEC}). We quote all parameter errors at the 1$\sigma$ (68\%) confidence level, assuming $\Delta$C-stat = 1.0 and $\Delta$C-stat=2.3 for single-- and two--parameter error estimates, respectively. When needed, we use the Akaike information criteria (AIC; \citealt{akaike1974new}) to investigate the significance of adding model components to the fit--function, which is calculated by $\Delta$AIC$ = -\Delta C +2\Delta k$ ($C$ is the C-stat and $k$ is the degree--of--freedom; \citealt{wen2018comparing}). The $\Delta$AIC $>$ 5 and $>$ 10 cases are considered a strong and very strong improvement, respectively, over the alternative model. For all the fits we perform in this paper, we include Galactic absorption using the model \texttt{TBabs} \citep{wilms2000absorption}. We fix the column density $N_{\rm H}$ to $5\times10^{20}{\rm cm}^{-2}$ without considering any intrinsic absorption, consistent with the work by \citep{margutti2019embedded,migliori2023roaring}. 

\subsection{\nus~observations}
AT2018cow was observed by \nus{} on five occasions since 2018-06-23 (MJD~58292). Due to the decreasing of the X--ray flux above 3~keV, AT2018cow was not detected in the last \nus{} observation (ObsID: 80502407002), and a count--rate upper limit of 1.1$\times10^{-4}$~counts~s$^{-1}$ has been inferred \citep{migliori2023roaring}. To perform spectral analysis, in this paper we only consider the first four \nus{} observations, which were all taken within 37~days since the discovery of the source. A list of \nus{} observations analyzed in this paper is presented in Table~\ref{tb:obslist}. We perform the \nus{} data reduction using {\sc NuSTARDAS} version 1.9.7 with calibration files updated on 2023-10-17 (version 20231017). We use the pipeline tool {\sc nupipeline} to extract the level--2 science data, and the tool {\sc nuproduct} to produce the source$+$background and background spectra from the level--2 data. For both FPMA and FPMB detectors onboard \nus{}, the source$+$background spectra are extracted from a circular source region of 30\arcsec{} radius centered on the source. The background spectra are extracted from circular apertures of $>$ 50\arcsec{} radii close to the source on the same detector, free from other bright sources.

\subsection{\xmm\ observations}
AT2018cow was observed by \xmm{} on three occasions within 300~days after its discovery, and another three occasions in the year 2022. Because the source becomes too faint that the background flux dominates over the source$+$background flux, we discard the last three \xmm{} observations (ObsID: 0843550401, 0843550501, 0843550601) in our subsequent analysis. We list the \xmm{} observations used for the analysis presented in this paper in Table~\ref{tb:obslist}. We reduce the \xmm{} data using HEASOFT version 6.32.1 and SAS version 21.0.0 with calibration files renewed on October 5th, 2023 (CCF release: XMM-CCF-REL-402). During one of the observations (ObsID: 0822580501), one of the two MOS detectors onboard \xmm{} was used for calibration and no science data was obtained. Meanwhile, the signal--to--noise ratio in the RGS detectors is too low to perform spectral analysis. Therefore, for consistency, we only use the data from the EPIC-pn detector. 

We use the SAS task {\sc epproc} to process the data. We employ the standard filtering criteria to exclude periods with an enhanced background count rate, requiring that the 10-12~keV detection rate of pattern 0 events is $<$ 0.4~counts~s$^{-1}$. We use a circular source region of 25\arcsec{} radius centered on the source for the source$+$background spectral extraction. This extraction region is somewhat smaller than we would have used normally be it is designed to avoid contamination from nearby soft--X-ray sources and the detector edges. Using the SAS command {\sc epatplot}, we check for the presence of photon pile--up and find no evidence for effects caused by pile--up. The background spectra are extracted from circular apertures of $\gtrsim$40\arcsec{} radii close to the source on the same detector, free from other bright sources. 

\subsection{\swf~observations}
In this paper we also include the X--ray data from the \swf{}/XRT instrument. \swf{} monitored AT2018cow within the first 100~days after its discovery. We extract the X--ray lightcurve from \swf{}/XRT using the online data reduction pipeline\footnote{https://www.swift.ac.uk/user$\_$objects/}\citep{evans2009methods}, applying the default reduction criteria. Meanwhile, using the same tool, we also extract the \swf{}/XRT source$+$background spectra as well as the background spectra (see \citealt{evans2009methods} for more details). Furthermore, for each of the \nus{} epoch, we combine the extracted \swf{}/XRT spectra that are taken on the same date. In this way, we prepare the quasi--simultaneous \swf{}/XRT observations for joint spectral analysis with the \nus{} observations. We present the information of these periods for the spectral count extraction in Table~\ref{tb:obslist}.

Throughout this paper, we carry out spectral analysis using the {\sc XSPEC} package (\citealt{arnaud1996xspec}; version 12.13.1). With the {\sc energies} command in {\sc XSPEC}, we create a logarithmic energy array of 1000 bins from 0.1 to 1000.0~keV for model calculations in all analyses for consistency.
Using the FTOOL {\sc ftgrouppha} for spectral analysis, we re--bin every background and source$+$background spectrum by the optimal--binning algorithm \citep{kaastra2016optimal}, while requiring the spectra to have a minimum of 1 count per bin (with parameter {\sc grouptype} in {\sc ftgrouppha} set to {\sc optmin}). For every spectrum, we discard the data bins where the background flux is dominating over the source$+$background flux. The remaining energy bands in each spectrum for our spectral analysis are given in Table~\ref{tb:obslist}. 

For every remaining observation, we first fit the background spectrum using a phenomenological model. When fitting the source$+$background spectrum, we then add the best--fit background model to the fit--function describing the source$+$background spectrum, with the background model parameters all fixed to their best--fit values determined from the fit to the background--only spectrum. The best--fit background model varies from instrument to instrument, and from epoch to epoch. For the \xmm{}EPIC-pn data, the best--fit phenomenological background model consists of between 2-3 power--law components and 2-3 Gaussian components; for \nus{}, it consists of 2 power--law and 3 Gaussian components; for \swf{}, it consists of 2 power--law components. The full--width at half--maximum (FWHM) of every background Gaussian component is fixed to $\sigma_{\rm Gauss}=0.001$~keV, less than the spectral resolution of either \xmm{}/EPIC-pn, \nus{}, or \swf{}. Such phenomenological models account for both the background continuum and the fluorescence lines \citep[e.g.,][]{katayama2004properties,pagani2007characterization,harrison2010nuclear}. In this paper, when studying the source$+$background spectra, we refer only to the part of the fit function that describes the source as fit function.

We group all the data mentioned above into six epochs by time of the observations. A list of the epochs and the associated observations can be found in Table~\ref{tb:obslist}. When performing spectral analysis, we always jointly fit the spectra within the same epoch using the same fit function for the source spectra. To account for the mission specific calibration differences, we use a constant component (\texttt{constant} in {\sc XSPEC}) multiplying the source models. This constant serves as a re--normalization factor between different instruments. Specifically, we fix the constant to be 1 for \nus{}/FPMA spectra, and let the constant for other instruments free--to--vary in the fits for each epoch.

\section{Extending slim disk model \texttt{slimdz} to lower $M_{\bullet}$}
\label{sc:model}

The very luminous X--ray emissions from AT2018cow (peak luminosity $L_x\gtrsim10^{44}$~erg~s$^{-1}$) imply the source is in the super--Eddington regime at least for the early days when its X--rays are near the peak, if powered by accretion onto a BH of $10^{1}-10^{3}$~$M_{\odot}$. When the mass accretion rate is at near--/super--Eddington levels, the accretion disk can no longer be adequately described by a standard thin disk model \citep{shakura1973black}, as the inward advection of the liberated energy is no longer negligible \citep[e.g.,][]{abramowicz1988slim}. Therefore, we choose to use a slim disk model, \texttt{slimdz} \citep{wen2022library}, to model the disk thermal emission from AT2018cow in our spectral analysis.

In its original form in \citealt{wen2022library}, 
the slim disk model
\texttt{slimdz} 
does not allow for {\sc XSPEC} fitting of BH masses $M_{\bullet}$
lower than 1000~$M_{\odot}$, because
the 
pre--calculated library of disk spectra
only extend to that BH mass limit.
To model the high--/super--
Eddington disk of a BH on $10^{1}-10^{3}$~$M_{\odot}$ mass scales, we modify 
\texttt{slimdz} by expanding the 
pre--calculated library 
down to 10~$M_{\odot}$. 
We follow the same procedures in \citealt{wen2022library} to calculate and ray--trace the disk spectrum given $M_{\bullet}$.

To have the new spectral library be consistent with the original library, we sample the $10^{1}-10^{3}$~$M_{\odot}$ mass range in the same way as the original $10^{3}-10^{5}$~$M_{\odot}$ mass range, simply scaling down all sampled values by two orders of magnitude. Then, for each sampled $M_{\bullet}$, we calculate the disk spectra for various mass accretion rates $\dot m$, inclinations $\theta$, and BH spins $a_{\bullet}$. We use the same sampled values of $\dot m$, $\theta$, and $a_{\bullet}$ used to construct the original library in \citealt{wen2022library}. 

Notably, when the disk is nearly face--on ($\theta<3^{\circ}$), the ray--tracing does not behave well, as described in \citealt{psaltis2011ray}. Thus, to avoid  errors and to keep the model self--consistent across different mass scales, we set the lower boundary of $\theta$ allowed in the modified \texttt{slimdz} to 3$^{\circ}$. This is slightly larger than the limit of 2$^{\circ}$ in the original \texttt{slimdz} model for more massive BHs. The reason is that, for the lower mass BH range we consider here, the curvature is larger, aggravating the problems with near--face--on ray--tracing.

We note that in \texttt{slimdz} model, the viscosity parameter $\alpha$ \citep{shakura1973black} is fixed to 0.1. Numerical simulations of super--Eddington accretion flows have shown that $\alpha\sim0.1$ for a BH of 10~$M_{\odot}$ and various $\dot m$ \citep{skadowski2015global}. Furthermore, calculations indicate that the impact of $\alpha$ (0.01--0.1) on the emergent spectrum is small compared to that of other model parameters like $M_{\bullet}$ or $a_{\bullet}$, across different $M_{\bullet}$ scales (from 10 to 10$^6$~$M_{\odot}$; e.g.,\citealt{dotan2011super,wen2020continuum}).

Meanwhile, although the disk radiative efficiency $\eta$ is fixed to 0.1 in the \texttt{slimdz} model, it is do so only for the purpose of determining the unit of the mass accretion rate $\dot m_{\rm Edd}=1.37\times10^{21}$~kg~s$^{-1}(0.1/\eta)(M_{\bullet}/10^{6}M_{\odot})$. The actual disk radiative efficiency can be determined from the physical value of $\dot m$ after constraining the mass $M_{\bullet}$, and the efficiency can vary between epochs (as expected in the slim disk scenario when the $\dot m$ changes; e.g., \citealt{abramowicz2013foundations}).

Moreover, \citealt{wen2022library} ray--trace the accretion disk only up to $\leq600$~$R_g$ (here $R_g=GM_{\bullet}/c^2$ is the gravitational radius of the BH), as regions with $R>600$~$R_g$ are expected to contribute little to the X--ray flux of TDEs due to the significant temperature decrease as a function of radius. \citet{wen2021mass} found that an error $\lesssim1\%$ in flux is introduced by this choice of the outer disk radius for the ray--tracing. However, TDEs with smaller mass BHs are likely to have larger disks ($\gtrsim2\times10^5$~$R_g$ for a 10~$M_{\odot}$ BH disrupting a solar--type star), and such disks are quantitatively different from those with $\geq10^4$~$M_{\odot}$ BHs. We test and find that, to keep the flux error $\lesssim1\%$ for most prograde--spinning BHs, we have to perform the ray--tracing further to at least 800~$R_g$ (Fig.~\ref{fig:nc}). In the Appendix, we show the comparison between the disk spectra with different choices of outer radii for the ray--tracing. Therefore, in our model calculations, the disk is ray--traced up to 800~$R_g$. We note that, in extreme cases of retrograde spinning BHs (e.g., $M_{\bullet}=10~M_{\odot}$, $a_{\bullet}=-0.998$, and $\dot m=100~\dot m_{\rm Edd}$), the flux errors introduced by this choice of the ray-tracing radius could be as large as $\sim$50\% $<$1~keV (Fig.~\ref{fig:ec}). One should therefore be cautious when interpreting the results in those cases.
 
Recently, by combining thin--disk results at the innermost--stable--circular--orbit (ISCO) with numerical simulations, \citealt{mummery2024continuum} finds that the disk plunging region inside the ISCO also contributes a significant part to the BH X--ray spectrum in some cases, associated with a finite stress at the ISCO due to a magnetic field and an extremely high spectral hardening factor of $f_c \sim 100$. In our case, the slim disk solution (e.g., temperature $T$ and surface density $\Sigma$ as a function of the disk radius $r$) describes the disk self--consistently from the event horizon to the outer disk edge \citep{skadowski2011relativistic,wen2020continuum}. When constructing the \texttt{slimdz} spectra library, we employ an inner boundary at the ISCO for the ray--tracing purpose only. \citealt{wen2021mass} have tested and found that the flux difference in the 0.2-10 keV energy band between ray--tracing down to the horizon and down to the ISCO is nearly always $\lesssim$2\%. Therefore, we conclude that the spectral impact of the disk plunging region is not significant in the slim disk cases of our interest, and for consistency we employ the ISCO as the inner boundary of the ray--tracing process in this paper as well.

\section{Results}

\begin{figure}
    \centering
    \includegraphics[width=\linewidth]{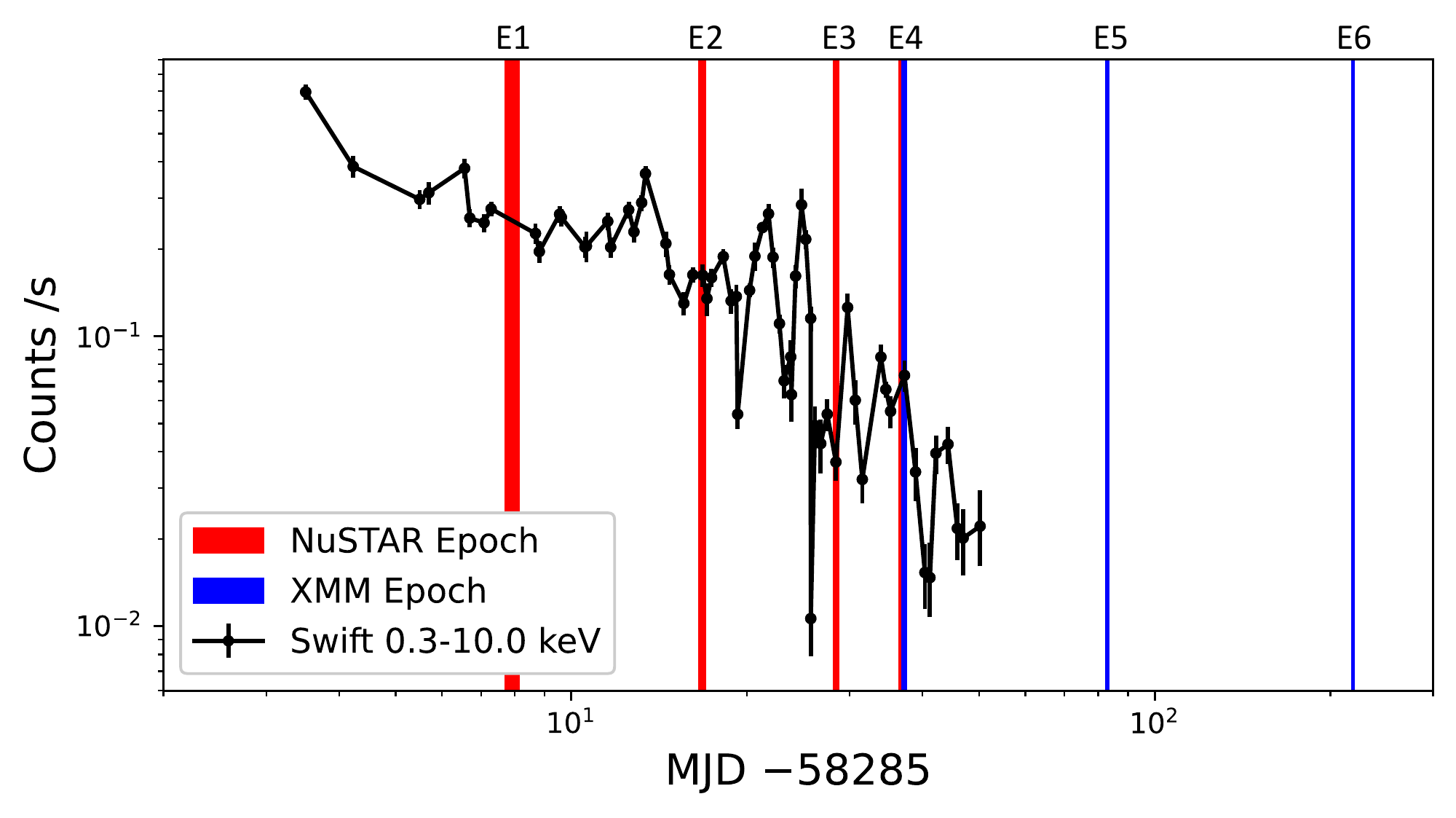}
    \caption{\swf{} 0.3--10~keV light curve of AT2018cow. The time of the \nus{} and \xmm{} observations has been highlighted with red and blue vertical lines, respectively. The 6 epochs we use in this paper for grouping the observations and joint analysis are marked at the top of the figure. See Table~\ref{tb:obslist} for the details of the observations in each epoch. Note at E4, the \nus{} and \xmm{} observations do not overlap in time.}
    \label{fig:lcs}
\end{figure}

Fig.~\ref{fig:lcs} shows the \swf{} 0.3-10.0~keV light curve, as well as the epochs of \nus{} and \xmm{} observations. We first explore the spectral characteristics by fitting the data with a fit function comprised of a power--law modified by the effects of Galactic extinction (the fit function in {\sc XSPEC}'s syntax is \texttt{"constant*TBabs*powerlaw"}). Results show that, except for the first epoch, spectra from the other epochs can be well--fit by a power--law (C-stat/d.o.f.$<$2; Table~\ref{tb:po}). From Epoch 2 to 6, the source becomes increasingly softer in X--rays, with the power--law index changing from $\Gamma=1.38\pm0.02$ (Epoch 2) to $2.8\pm0.6$ (Epoch 6). The soft source spectrum at the latter epoch might indicate the appearance of a soft disk component in the energy range 0.3-1.5~keV. 

\begin{figure}
    \centering
    \includegraphics[width=\linewidth]{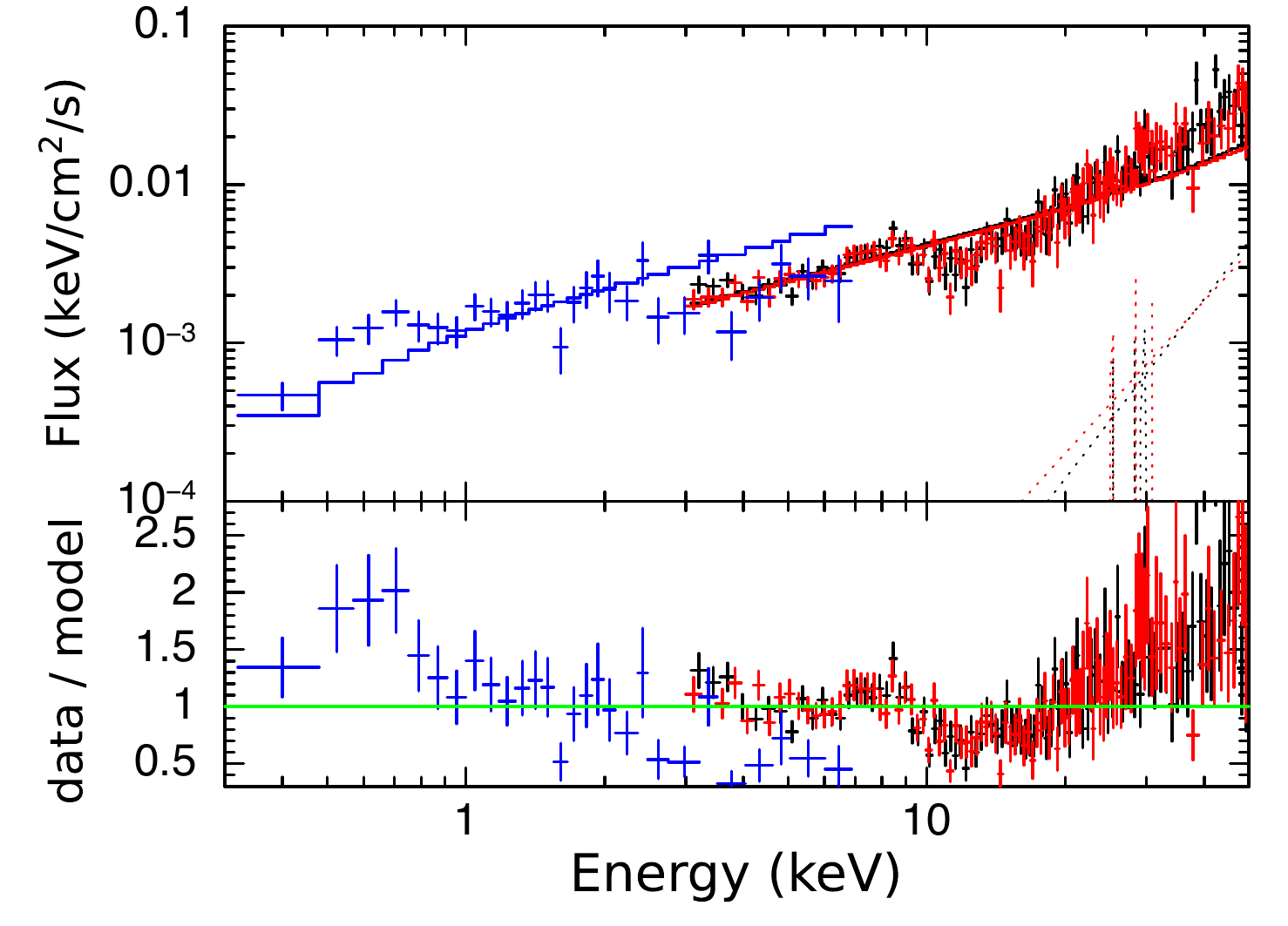}
    \caption{The X--ray spectra from Epoch 1 fitted by a power--law model. In the upper panel we present the data, the power--law model (solid lines), and the background models (dotted lines). The blue, black, and red data are from \swf{}/XRT, \nus{}/FPMA, and \nus{}/FPMB, respectively. In the lower panel, we show the ratio between the observed number of counts (data) and the predicted number of counts in each spectral bin (model). Similar to what previous studies find, we observe X--ray features around $\sim$6.4~keV and above 10~keV that are likely due to reflection.}
    \label{fig:1e-po-res}
\end{figure}

From the fit residuals for Epoch 1 (Fig.~\ref{fig:1e-po-res}), we confirm the X--ray features $\sim$6.4~keV and $\gtrsim$10~keV as found previously \citep{margutti2019embedded}. These features cause the source spectrum to be inconsistent with a single power--law fit function at Epoch 1, and they have been proposed to be due to the reflection of the primary power--law emission (possibly caused by a black hole corona or a jet base) off an accretion disk. 

We then use the model \texttt{relxillCp} (\citealt{dauser2014role,garcia2014improved}; see also the references for a detailed description of each model parameter) to account for the disk reflection in the fit function. We note, in \texttt{relxillCp}, the disk is assumed to be a standard Shakura--Sunyaev thin disk \citep[][]{shakura1973black}, and the incident power--law emission is modeled by \texttt{nthcomp} \citep{zdziarski1996broad}, that assumes a multi--temperature black body seed spectrum modified by a Comptonising medium. Currently, there are no reflection models using a slim disk for the disk seed photons or calculating the GR effect, and so we use \texttt{relxillCp} to approximate the reflected emission off a slim disk (the total fit function in {\sc XSPEC}'s syntax is \texttt{"constant*TBabs*relxillCp"}). Therefore, one should be cautious when interpreting the results from our fits with \texttt{relxillCp}. At Epoch 1, we find the source emission is consistent with \texttt{relxillCp}, i.e.~an incident power--law--like emission plus a disk reflection (Table~\ref{tb:1to6}). The Comptonising medium is constrained to be emitting a hard power--law ($\Gamma=1.22\pm0.02$), and its electron temperature is constrained to be $28\pm8$~keV. The inclination constraint is $74^{\circ}\pm2$ and the black hole spin constraint is $0.98\pm0.01$.

We then test for the presence of a spectral component originating from an accretion disk in the other epochs of AT2018cow, using 
\texttt{slimdz}. Results are summarized in Table~\ref{tb:1to6}. Among Epoch 2 to 6, we find that only for Epoch 4 the fit to the data is significantly improved by adding a disk component to the fit function (the total fit function in {\sc XSPEC}'s syntax is \texttt{"constant*TBabs*(powerlaw+slimdz)"}; $\Delta$AIC=12.8 compared to the power--law--only case, exceeds the very strong improvement threshold of 10; \citealt{akaike1974new}); statistically, Epoch 2, 3, 5, and 6 require no disk components besides a simple power--law for reaching a good fit to the data. Meanwhile, a slim disk alone (\texttt{"constant*TBabs*slimdz"} in {\sc XSPEC}'s syntax) cannot describe the data well, for Epoch 2, 3, and 5. However, as the source becomes much softer at Epoch 6, we test and find that the source spectrum at this epoch is consistent with the slim disk model, yielding a black hole mass log($M_{\bullet}/M_{\odot}$)=$2.3\pm0.9$, while due to data quality, other disk parameters are not well constrained ($\dot m$, $\theta$, $a_{\bullet}$). The constraints on the black hole mass derived from the spectra both at Epoch 4 and at Epoch 6 are consistent with log($M_{\bullet}/M_{\odot}$)$\approx2.4$.

\begin{figure*}
    \centering
    \includegraphics[width=0.5\textwidth]{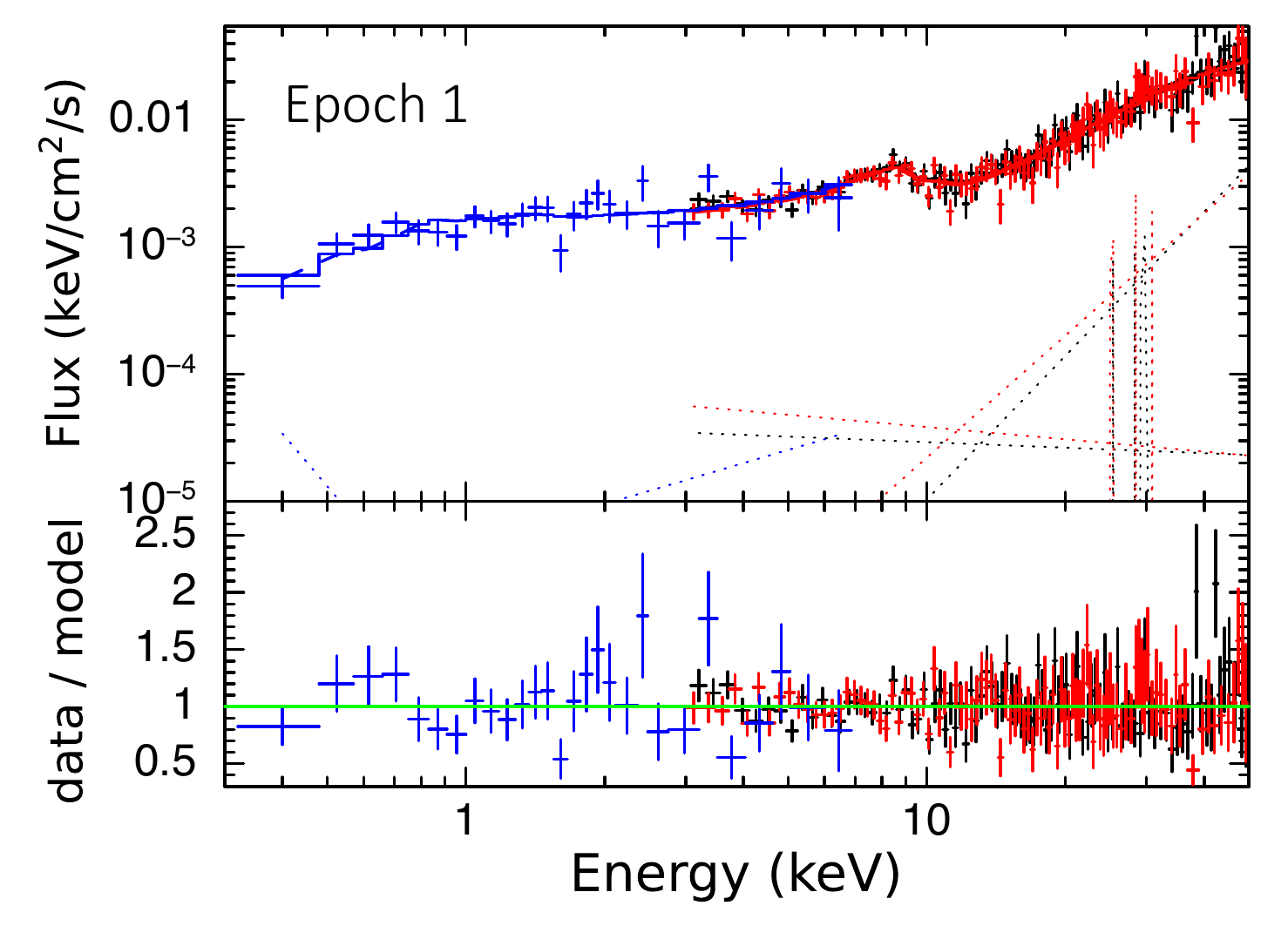}\hfill
    \includegraphics[width=0.5\textwidth]{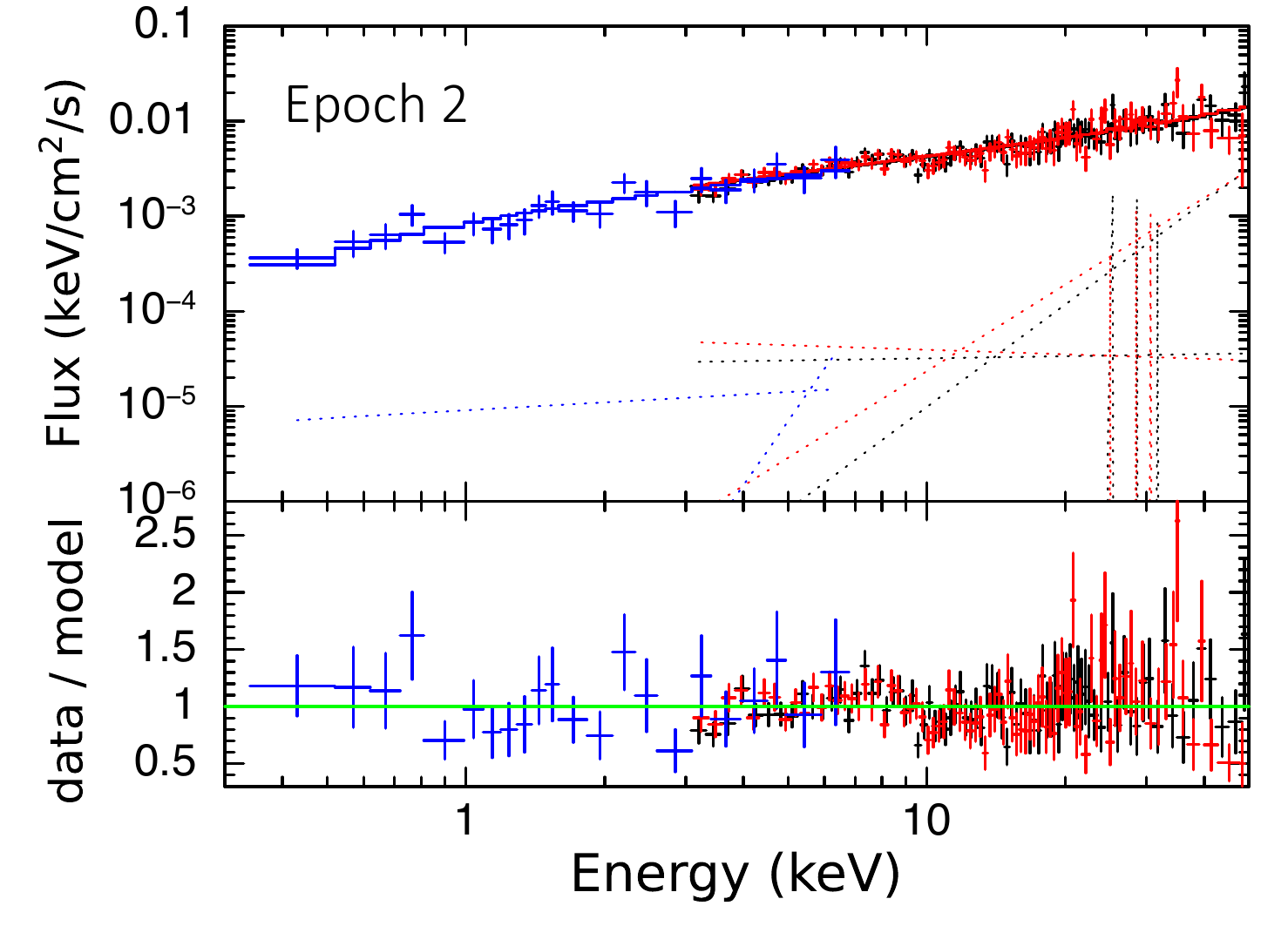}\hfill
    \includegraphics[width=0.5\textwidth]{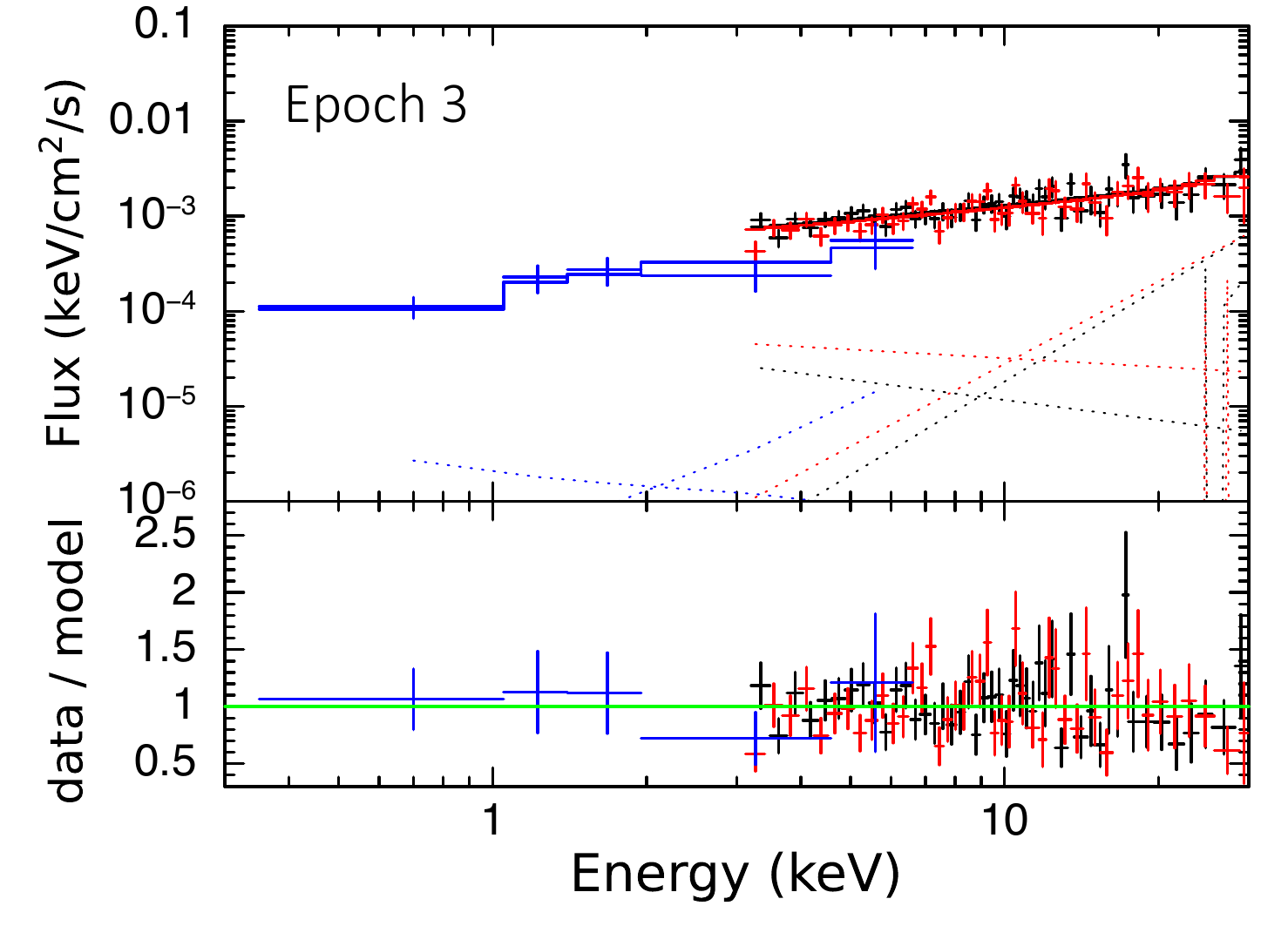}\hfill
    \includegraphics[width=0.5\textwidth]{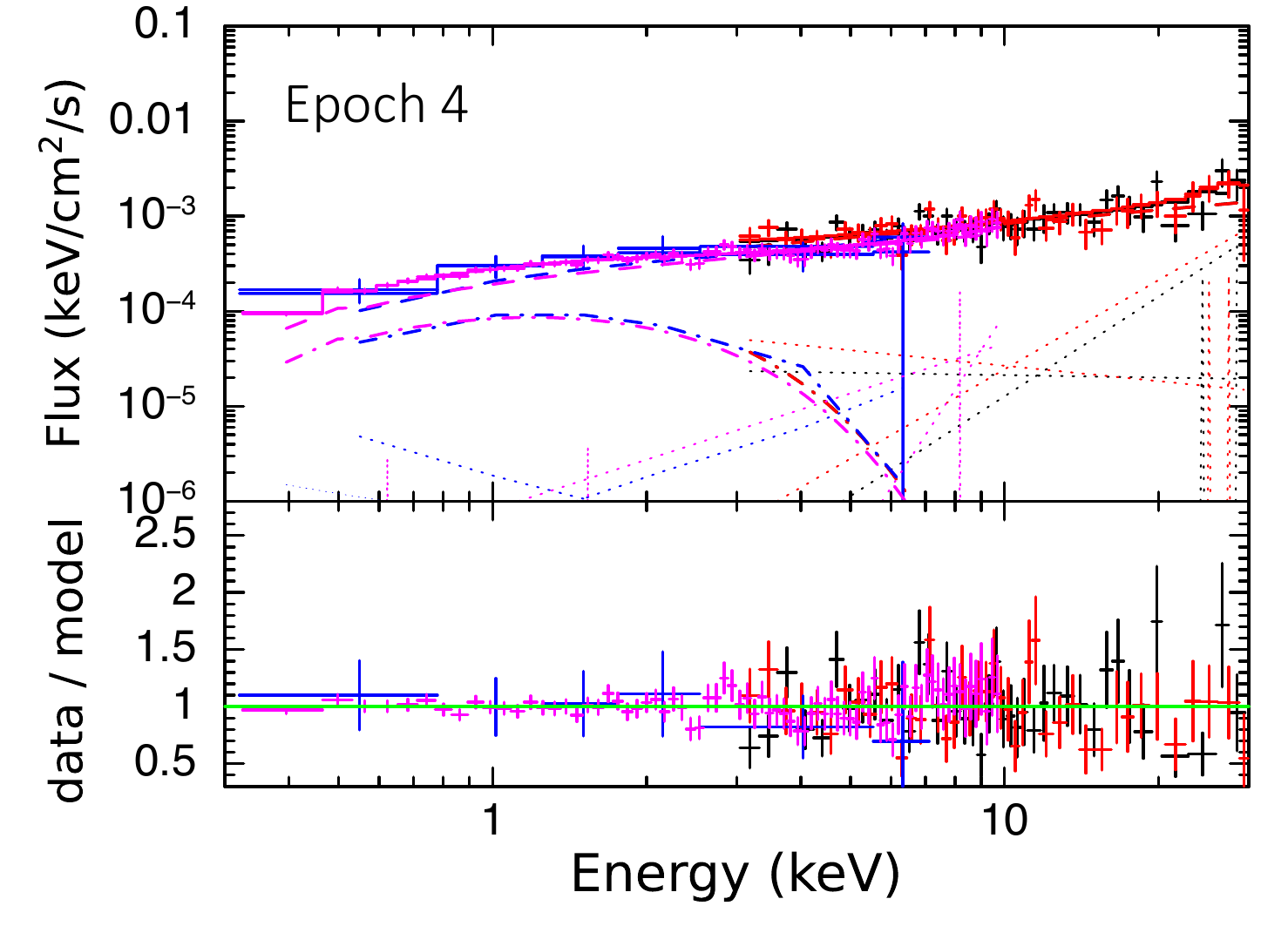}\hfill
    \includegraphics[width=0.5\textwidth]{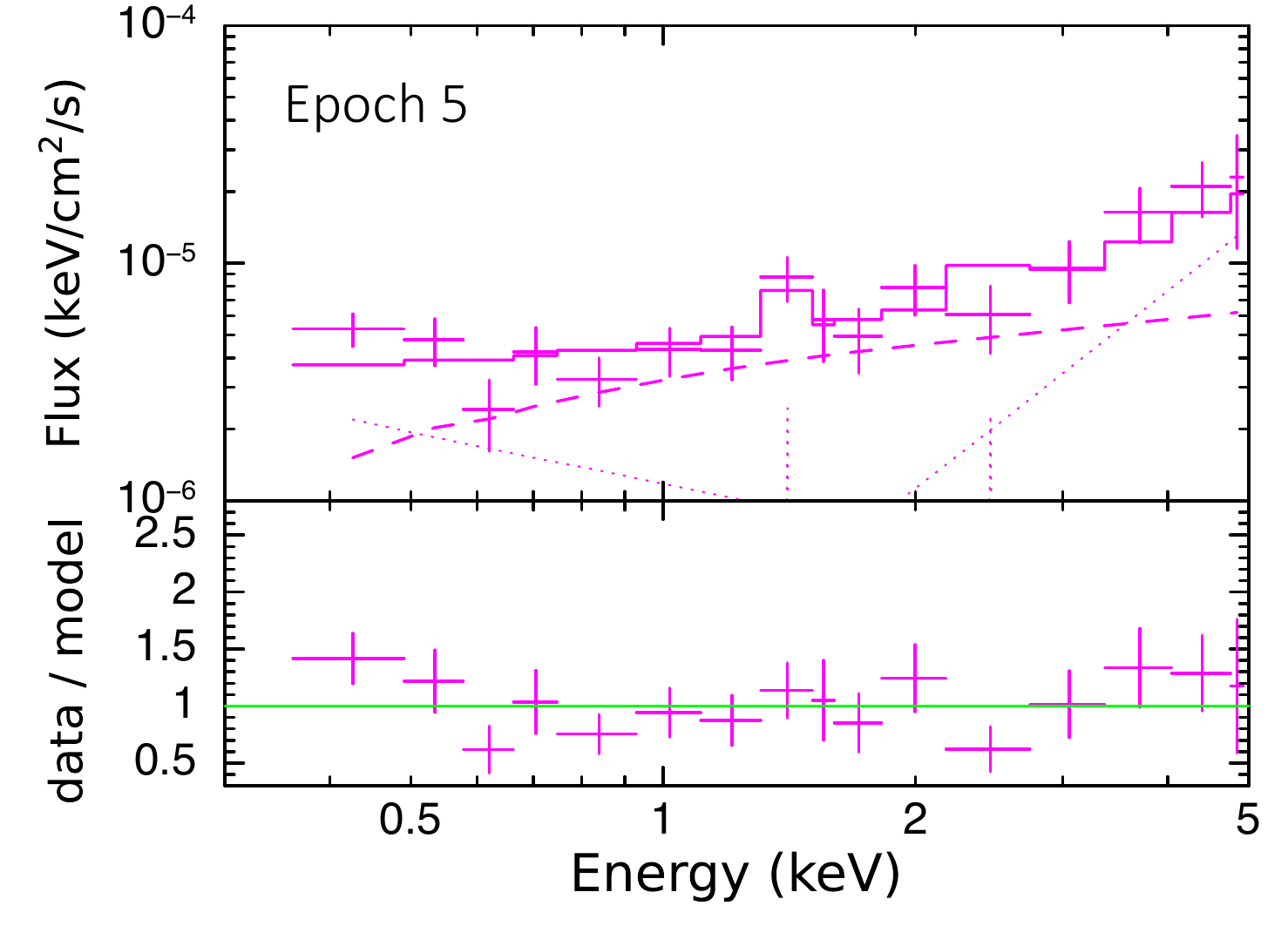}\hfill
    \includegraphics[width=0.5\textwidth]{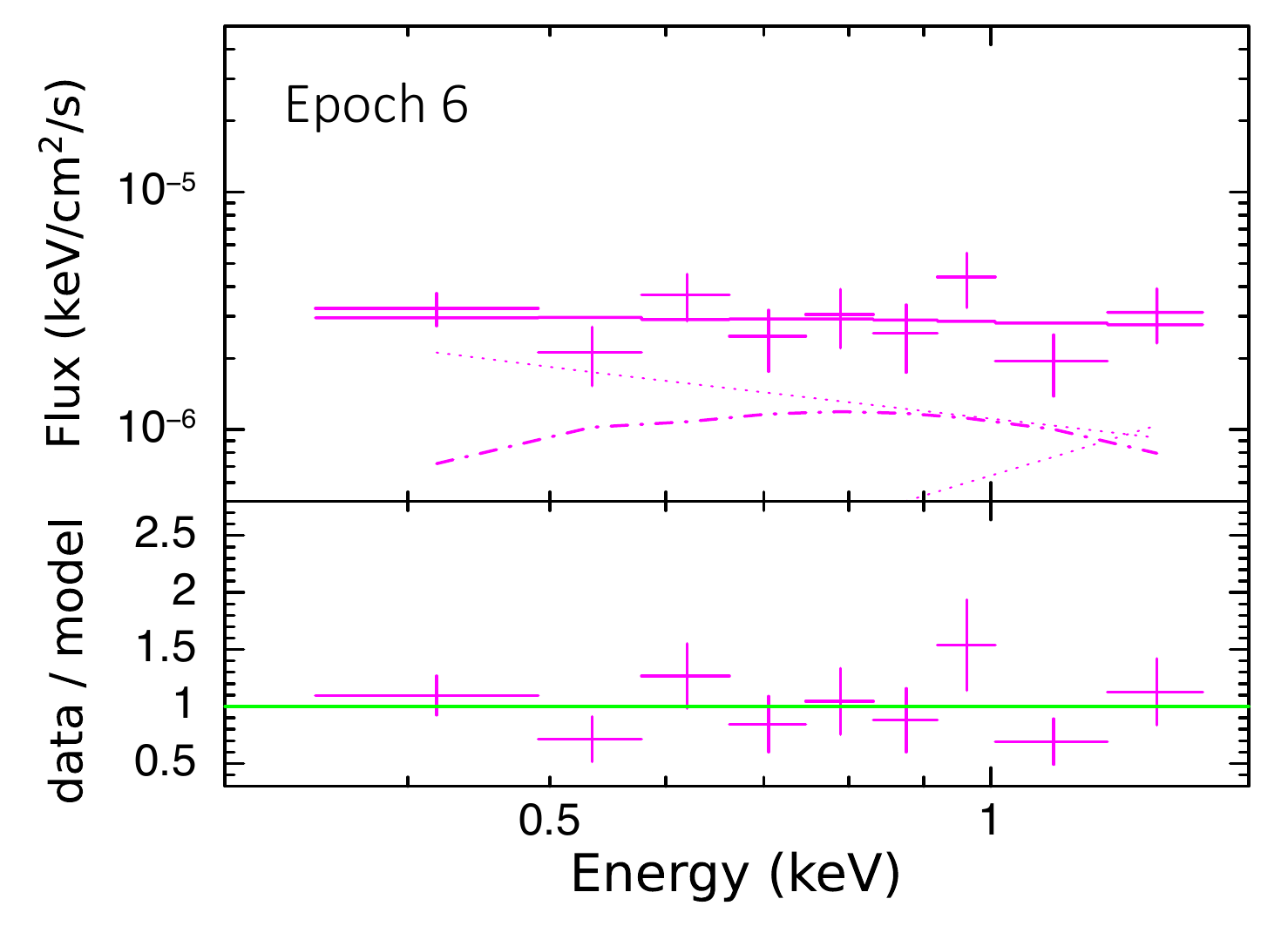}
    \caption{The best--fit results based on the spectral analysis presented in Table~\ref{tb:1to6}. For each figure, in the upper panel we present the spectrum as well as the best--fit models, and in the lower panel we present the ratio between the observed number of counts (data) and the best--fit predicted number of counts in each spectral bin (model). In all figures, the solid, dashed, dot--dashed, and dotted lines represent the best--fit source$+$background model, the power--law spectral component, the slim disk component, and the background models, respectively; the blue, black, red, and magenta data are from \swf{}/XRT, \nus{}/FPMA, \nus{}/FPMB, and \xmm{}/EPIC-pn, respectively. Notice the y--axes in the upper panels of each figure are different in scales for illustration purpose.}
    \label{fig:1to6}
\end{figure*}

Physically, it is likely that the accretion disk is present around the black hole not only at Epoch 1, 4, and 6, but also throughout the period (weeks) after the first detection (Epoch 1). At Epoch 2 and 3, the soft X--ray band (0.3--3.0~keV) can only be investigated through the \swf{}/XRT data. As the slim disk emits primarily soft X--rays, a lack of data of higher quality in the soft band, together with the presence of a strong power--law emission component, makes it impossible to constrain the disk emission at Epoch 2 and 3. At Epoch 4, however, the high quality 0.3--3.0~keV \xmm{} data allows a significant measurement of this soft disk component. At Epoch 5, the general decrease in the source luminosity makes it impossible to detect the disk in the \xmm{} observation especially since the source spectrum remains dominated by a power--law. At Epoch 6, although the luminosity keeps decreasing, we find the source spectrum has become much softer (Table~\ref{tb:po}). It is likely that at this epoch the luminosity of the non-thermal component has diminished and the spectrum can be explained solely by the disk emission. The disk model fits the data well without the need of non-thermal components (Fig.~\ref{fig:1to6}).

\begin{figure}
    \centering
    \includegraphics[width=\linewidth]{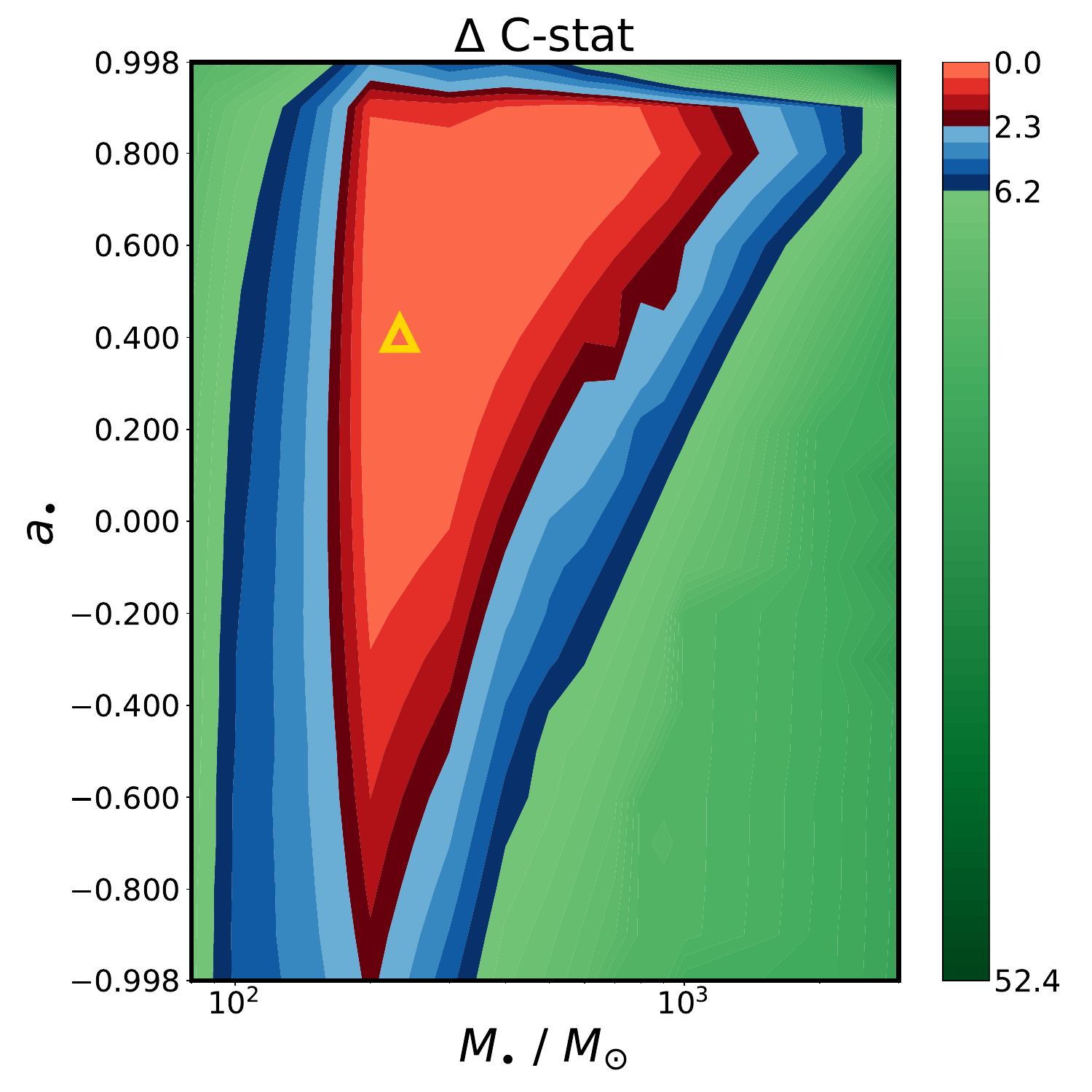}
    \caption{Constraints on $M_{\bullet}$ and $a_{\bullet}$ based on a joint fit of the X-ray spectra from Epoch 2 to 6 with the slim disk model (Table~\ref{tb:jointfit}). We calculate the $\Delta$C-stat across the $\{M_{\bullet}$, $a_{\bullet}\}$ plane. The best--fit point with the lowest C-stat is marked by the yellow triangle. Areas within 1$\sigma$ and 2$\sigma$ confidence levels are filled by red and blue colours, respectively. At 1$\sigma$ for the two--parameter fits, $M_{\bullet}$ is constrained to be log($M_{\bullet}/M_{\odot}$)$=2.4^{+0.6}_{-0.1}$, while $a_{\bullet}$ is virtually unconstrained.}
    \label{fig:contour}
\end{figure}

Therefore, based on the results above, we assume the disk is present at all epochs and perform joint--fits combining the data from Epoch 2 to 6. The total fit function in {\sc XSPEC}'s syntax is \texttt{"constant*TBabs*(powerlaw+slimdz)"}. We force the disk parameter $\theta$, $M_{\bullet}$ and $a_{\bullet}$ to be the same between epochs and fit their values, while we allow $\dot m$ to vary between epochs and treat each $\dot m$ as an individual fit parameter for each epoch. The power--law emission is also allowed to vary between epochs in our joint--fit. By jointly fitting Epoch 2 to 6, we find the black hole mass to be log($M_{\bullet}/M_{\odot}$)$=2.4_{-0.1}^{+0.6}$, an upper limit to the inclination $\theta<76^{\circ}$, and a broadly--constrained black hole spin $a_{\bullet}=0.4^{+0.5}_{-0.9}$. We present the full list of parameter constraints from the joint--fit in Table~\ref{tb:jointfit}, and the $\Delta$C-stat contour in $\{M_{\bullet}$, $a_{\bullet}\}$ space in Fig.~\ref{fig:contour}. 

The data at Epoch 4 and 6 plays a vital role in constraining the value of the BH mass. Nonetheless, it is important that we also consider the data at other epochs, because they will help us constrain parameters that might vary over the event and it can help us exclude models that are not consistent with the data (e.g., upper limits on the disk luminosity can be derived). Notably, the inclination constraint derived from the thin--disk reflection model at Epoch 1 does not deviate from the slim disk results based on this joint--fit, while the thin--disk reflection model suggests a higher black hole spin than the slim disk results. 
The $\dot m$ constraints are consistent with a scenario in which the accretion rate decreases from super--Eddington to sub--Eddington levels. 

In all the fits above, we notice the re--normalization constant for \swf{} spectrum at Epoch 3 is constrained to be $\sim$0.5, which stands out from the re--normalization of \swf{} observations at other epochs. We manually check the CCD image for this particular \swf{} observation and find a stripe of dead pixels to be present in the source extraction region. We also checked the other \swf{} data and find this row of dead pixels to lie away from the source extraction region. This defect on the CCD leads to a loss in the instrumental effective area and thus results in a decrease of the number of counts in this particular spectrum, explaining the lower constant in the joint--fit of the \swf{} spectra with those of the other satellite data.

\section{Discussion} 
\label{sc:discuss}

By analyzing AT2018cow data from \swf{}, \nus{}, and \xmm{}, we find that starting from Epoch 2 the source's X--ray spectrum can be interpreted as originating from a slim disk plus a non--thermal spectral component modeled by a power--law. For the X-ray spectral fits, we extend the black hole mass range available for the slim disk model \texttt{slimdz} \citep{wen2022library} from $10^3-10^6$~M$_\odot$ to the black hole mass range $10^1-10^6$~M$_\odot$. This extension to the model \texttt{slimdz} is now publicly available\footnote{10.5281/zenodo.11110331} to be used in {\sc XSPEC}. We confirm that AT2018cow's X-ray spectrum during Epoch 1 can be well-described by a disk reflection model as was reported before (\citealt{margutti2019embedded}). When the source becomes softer in X--rays after $\gtrsim$200~days, the source spectrum becomes consistent with the emission from a slim disk. From the slim disk modeling, an IMBH of mass log($M_{\bullet}/M_{\odot}$)$=2.4^{+0.6}_{-0.1}$ (at 68\% confidence level) is derived for the mass of the central compact object, while the disk inclination ($\theta<76^{\circ}$) and the BH spin ($a_{\bullet}=0.4^{+0.5}_{-0.9}$) are less strongly constrained (see Table~\ref{tb:jointfit} for all the parameter constraints). All the parameter constraints are derived under the assumption that the disk viscosity parameter $\alpha=0.1$. Our spectral modeling shows that the X--ray spectrum becomes softer at late time due to the disappearance of the non--thermal spectral component in concert with a decrease in the mass accretion rate, $\dot m$. The $\dot m$ values derived from our spectral fit are consistent with a decrease from super-- to sub--Eddington levels at late--times.

Our independent mass measurement of AT2018cow is consistent with several mass constraints available in the literature based on energetic arguments, late--time UV data modeling, and  X--ray timing assumptions \citep{pasham2022evidence,migliori2023roaring,inkenhaag2023late}. The mass constraint log($M_{\bullet}/M_{\odot}$)$=2.4^{+0.6}_{-0.1}$ confirms the presence of an accreting IMBH as the central compact object. Such an IMBH can be formed through the accretion of gas onto a seed stellar--mass BH (which can take cosmic timescales; e.g., \citealt{madau2001massive,greif2011simulations}), or through the direct collapse of pristine gas clouds in the early Universe \citep[e.g.,][]{loeb1994collapse,bromm2003formation,lodato2006supermassive}, or through the BH merger events \citep[e.g.,][]{abbott2016properties}. The direct mass measurement from the slim disk modeling demonstrates a possible new way to study other AT2018cow--like LFBOTs that are also accompanied by variable X--ray emission (e.g., AT2020xnd/ZTF20acigmel, \citealt{ho2022luminous,bright2022radio}; AT2020rmf, \citealt{yao2022x}). 

An IMBH nature for AT2018cow is in line with scenarios that involve an accreting central compact object (e.g., TDE; \citealt{kuin2019swift,perley2019fast}; BH--star binary merger; \citealt{metzger2022luminous}). Particularly, for the IMBH--TDE scenario, we estimate the late--time disk outer radius to be $\approx13$~$R_{\odot}$ given an IMBH of mass $10^{2.4}\approx250$~$M_{\odot}$ disrupting a solar--mass star\footnote{While the \texttt{slimdz} model limits the disk outer radius to $\leq800~R_g$ (for ray-tracing purposes; see  Section~\ref{sc:model}), in the case of AT2018cow, we estimate an error on the flux of $\lesssim0.5\%$ introduced to the model below 1~keV by this choice of disk radius (Fig.~\ref{fig:oc}). Meanwhile, the actual disk radius is estimated by assuming a typical TDE disk, whose outer radius is about twice the tidal radius: $R_{\rm out}\approx2R_t=2R_{\rm star}(M_{\bullet}/M_{\rm star})^{1/3}$ \citep[e.g.,][]{rees1988tidal,kochanek1994aftermath}. See also the Appendix.}. Thus, the predicted radius from the IMBH--TDE scenario is similar to that from the binary merger scenario (15$-$40~$R_{\odot}$; \citealt{migliori2023roaring}) and has the same order of magnitude as values derived from the late--time UV observations ($\approx40$~$R_{\odot}$; \citealt{inkenhaag2023late,migliori2023roaring}). Meanwhile, in the core--collapse scenario, the fall--back stellar ejecta typically forms a much smaller disk, and this generally leads to an outer disk radius of $\sim10^{-3}$~$R_{\odot}$ at late times \citep{migliori2023roaring}. Besides the outer disk radius, other parameters might also differ among scenarios, e.g., the peak $\dot m$. We refer to previous studies for a quantitative discussion of how those parameters depend on the BH mass in different scenarios \citep[e.g.,][]{metzger2022luminous,migliori2023roaring}. 

In our study, the non--thermal spectral component is modeled by a power--law component. Possible origins of the power--law component include a Comptonising medium (BH corona) up--scattering the disk photons, similar to that found in other BH accretion systems like X--ray binaries or Active Galactic Nuclei (AGNs) \citep[e.g.,][]{esin1997advection,nowak2011corona}. It is also possible that some of the non--thermal emission is generated by shock interactions with the circumstellar material (CSM; e.g., \citealt{rivera2018x,margutti2019embedded,fox2019signatures,leung2020model}. 
Interestingly, when the accretion rate becomes sub--Eddington at late times, we find the non--thermal emission diminishes and the X-ray spectrum can be well-fit by a slim disk model. Spectral state transitions involving a varying non--thermal spectral component has also been observed in several TDE systems \citep[e.g.,][]{bade1996detection,komossa2004huge,wevers2019black,jonker2020implications,wevers2021rapid,cao2023rapidly}.

In AT2018cow, there is evidence for a dense CSM (\citealt{ho2019at2018cow}). Theoretical work shows that a dense CSM can be present in both a TDE scenario (\citealt{linial2024tidal}), where the dense CSM is produced by the outflows from the BH--star mass transfer prior to the full stellar disruption, and in a binary merger scenario, 
where the dense CSM is produced in the common envelope phase between a stellar--mass BH (1--20~$M_{\odot}$) and its massive stellar companion \citep{metzger2022luminous}. At present, it is unclear if this latter model can be extrapolated to accommodate a BH of $\approx250~M_{\odot}$, the value suggested by our work.  If that extrapolation  is plausible, our BH mass determination of AT2018cow is consistent with both the TDE and BH--star merger scenarios.

Since no reflection models for the reflected emission from a slim disk are available, in our analysis, the reflection features (i.e., the broadened iron K$\alpha$ line $\sim$6.4~keV and the Compton hump $>$10~keV) dominating the first \nus{} epoch (and not detected in all later epochs) are modeled by disk reflection \texttt{relxillCp} \citep{dauser2014role,garcia2014improved}. The model \texttt{relxillCp} assumes that a standard thin disk reflects the emission from a Comptonising medium with an incident power--law spectral shape. Given the same Comptonising medium, the geometrical differences between the thin and the slim disks will result in differences in the emissivity profile of the reflected emission. This inconsistency between the likely super--Eddington accretion and the thin--disk assumption at Epoch 1 might contribute to the different spin constraints derived from the thin--disk ($a_{\bullet}=0.98\pm0.01$) and the slim--disk ($a_{\bullet}=0.4^{+0.5}_{-0.9}$) results. Despite that, we notice the inclination constraint from \texttt{relxillCp} ($\theta=74^{\circ}\pm2$) is in general agreement with the value derived from the slim disk modeling using data from all later epochs ($<76^{\circ}$). The material that is responsible for the reflected emission can involve the rapidly expanding outflow. Its density will decrease with time, which leads to the diminishing of reflection features in later epochs \citep[e.g.,][]{margutti2019embedded}. 

While we extend the pre--calculated library of the disk spectra in \texttt{slimdz} to model the disk of $M_{\bullet}<1000$~$M_{\odot}$, there exists a slim disk model \texttt{slimbh} \citep{skadowski2011relativistic,straub2011testing} that is available for the disk luminosity $L_{\rm disk}\leq1.25$~$L_{\rm Edd}$ (with $L_{\rm disk}$ as one of the fit parameters in \texttt{slimbh}, and $L_{\rm Edd}\equiv1.26\times10^{38}(M/M_{\odot})$~erg/s)\footnote{Note in \texttt{slimdz}, instead of $L_{\rm disk}$, $\dot m$ in the unit of $\dot M_{\rm Edd}\equiv1.37\times10^{15}(M/M_{\odot})$~kg/s is used to solve the disk equations that determines the disk luminosity in erg/s.}. Physically, compared to \texttt{slimbh}, the \texttt{slimdz} model includes the loss of angular momentum due to radiation at each disk annulus. This adjustment alters the predicted effective temperature of the inner disk region, especially for high-spin, low-accretion disks \citep{wen2021mass}. Moreover, a different estimate of the disk spectral hardening factor $f_c$ \citep{davis2019spectral} is employed by \texttt{slimdz} compared with \texttt{slimbh}.

We compare the results derived by using \texttt{slimdz} with those obtained using \texttt{slimbh} by jointly fitting Epoch 2 to 6 with the fit function \texttt{"constant*TBabs*(powerlaw+slimbh)"}, and comparing the results (presented in Table~\ref{tb:slimbh}) to the those obtained with \texttt{slimdz} (Table~\ref{tb:jointfit}). Both disk models provide a good fit to the combined data from Epoch 2 to 6. The mass constraint derived from \texttt{slimbh} is slightly higher (log($M_{\bullet}/M_{\odot}$)$=3.2\pm0.2$) although marginally consistent with that in the \texttt{slimdz} case (log($M_{\bullet}/M_{\odot}$)$=2.4^{+0.6}_{-0.1}$). Besides the physical differences between the disk models, it is possible that when fitting the super--Eddington spectra at Epoch 4, the upper parameter range of $L_{\rm disk}/L_{\rm Edd}\leq1.25$ in \texttt{slimbh} limits the lower boundary of the mass constraint (since for a given observed luminosity, $L_{\rm disk}/L_{\rm Edd}\propto1/M_{\bullet}$). In this sense, as $L_{\rm disk}/L_{\rm Edd}$ at Epoch 4 is constrained to be $>$1.0 and is limited by the parameter range of $\leq$1.25, the lower boundary of the mass constraint derived from \texttt{slimbh} is underestimated, causing the mass constraints between the two disk models to be just marginally consistent in Table~\ref{tb:jointfit} and Table~\ref{tb:slimbh}. Indeed, the mass constraints derived from both disk models are consistent when jointly fitting Epoch 5 and 6 (Table~\ref{tb:last2}), at which two epochs the disk is likely at sub--Eddington accretion levels. We also test the $\alpha=0.01$ cases in the \texttt{slimbh} model. The fit results with $\alpha=0.01$ do not differ from those with $\alpha=0.1$ shown in Table~\ref{tb:slimbh}~and~\ref{tb:last2} (for \texttt{slimbh} cases). This test suggests that the choice of $\alpha=0.1$ does not significantly affect the slim disk spectrum or the BH mass constraint, though the spectral library for \texttt{slimdz} does not currently extends to lower $\alpha$ values. For the above reason concerning the upper limit of $L_{\rm disk}/L_{\rm Edd}$, and due to the improved treatment of the angular momentum transport by radiation \texttt{slimdz} mentioned above, we prefer the results derived from \texttt{slimdz} (Table~\ref{tb:jointfit}).

\section{Conclusions}

We have performed X--ray spectral analysis on \nus{}, \swf{}, and \xmm{} data of the LFBOT AT2018cow and find evidence for an accretion disk soon after the source's discovery. Based on slim disk modeling, we constrain the mass of the central compact object to be log($M_{\bullet}/M_{\odot}$)$=2.4^{+0.6}_{-0.1}$ at the 68\% confidence level. Our mass measurement is independent from, but consistent with, the results from previously employed methods. Therefore, we provide further evidence for an accreting intermediate--mass black hole ($10^{2}$---$10^{6}~M_{\odot}$) as the central compact object residing in AT2018cow, and by extension, in similar LFBOT sources. The mass constraint is consistent with both the tidal disruption and the black hole--star merger scenarios, if the latter model can be extrapolated to the measured black hole mass.

Our results are consistent with the scenario in which the source accretion rate decreases from super--Eddington to sub--Eddington levels $\sim$200~days after its discovery. We find the late--time spectrum to be softer compared to the early--time spectra, consistent with emission from a slim disk at sub--Eddington accretion levels. In our analysis, a modified version of the existing slim disk model \texttt{slimdz} is used to model the high--/super-- Eddington disk of a black hole at mass scales of 10 to 1000 $M_{\odot}$. Through this work, we demonstrate a possible new way of studying LFBOT sources that have X--ray emission similar to AT2018cow.

\begin{acknowledgements}
We thank the referee for comments that helped to improve this manuscript. This work used the Dutch national e-infrastructure with the support of the SURF Cooperative using grant no.~EINF-3954. This work made use of data supplied by the UK Swift Science Data Centre at the
University of Leicester. P.G.J.~has received funding from the European Research Council (ERC) under the European Union’s Horizon 2020 research and innovation programme (Grant agreement No.~101095973).
AIZ acknowledges support in part from grant NASA ADAP \#80NSSC21K0988 and grant NSF PHY-2309135 to the Kavli Institute for Theoretical Physics (KITP). 
\end{acknowledgements}

\section*{Data Availability}

All the X--ray data in this paper are publicly available from the HEASARC data archive (https://heasarc.gsfc.nasa.gov/). A reproduction package is available at DOI: 10.5281/zenodo.11110331. The extension to the \texttt{slimdz} model used in this paper is available at 10.5281/zenodo.11110331.

\bibliographystyle{aa}
\bibliography{reference}  

\newpage
\appendix

\section{The choice of the outer disk radius in \texttt{slimdz}}

\begin{figure}
    \centering
    \subfloat[\label{fig:nc}$M_{\bullet}=10~M_{\odot},~a_{\bullet}=0,~\dot m=\dot m_{\rm Edd}$, $\theta=45^{\circ}$]{\includegraphics[width=0.9\linewidth]{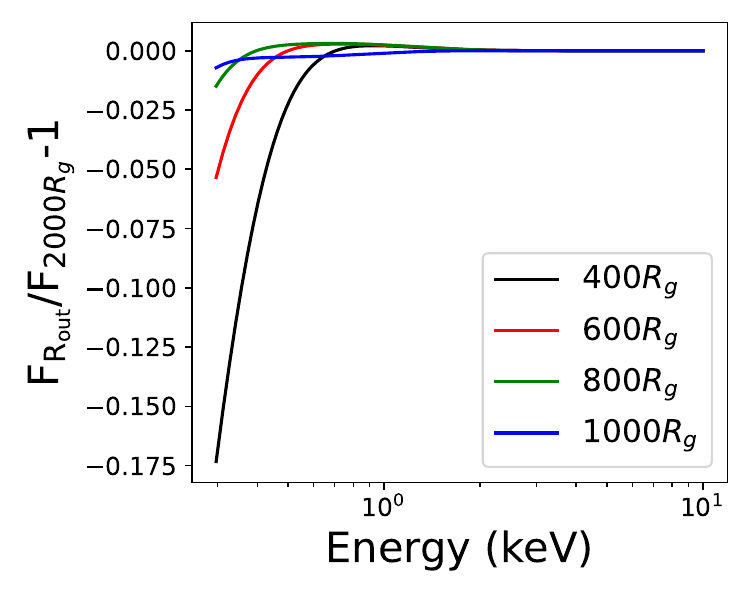}}\hfill
    \subfloat[\label{fig:ec}$M_{\bullet}=10~M_{\odot},~a_{\bullet}=-0.998,~\dot m=100~\dot m_{\rm Edd}$, $\theta=45^{\circ}$]{\includegraphics[width=0.9\linewidth]{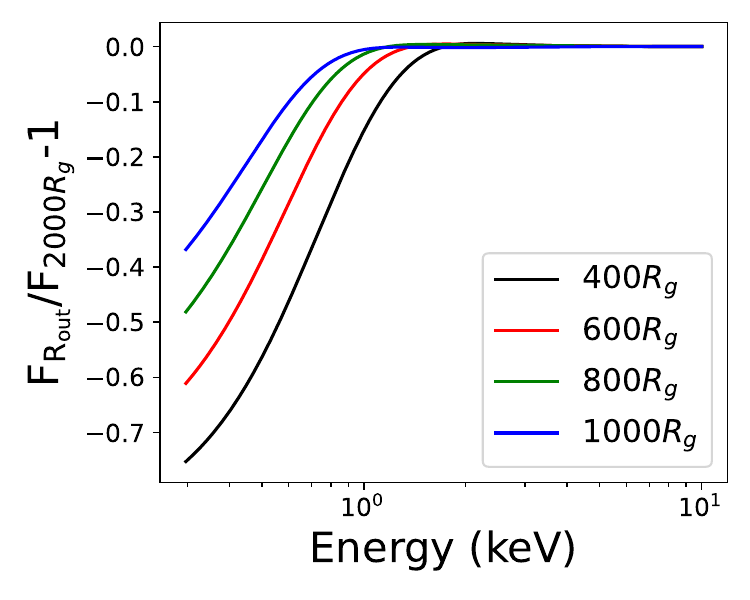}}\hfill
    \subfloat[\label{fig:oc}$M_{\bullet}=250~M_{\odot},~a_{\bullet}=0.4,~\dot m=5~\dot m_{\rm Edd}$, $\theta=74^{\circ}$]{\includegraphics[width=0.9\linewidth]{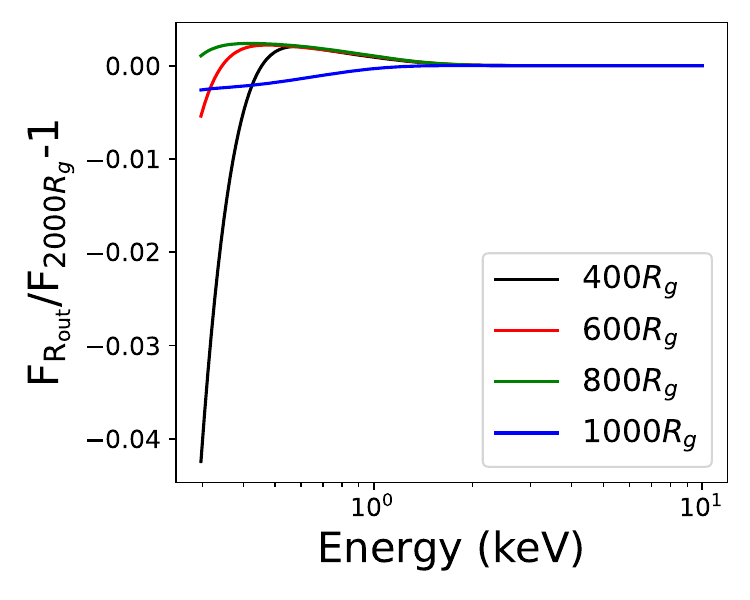}}\hfill
    \caption{Here we investigate the effect of different choice of the outer disk radius $R_{\rm out}$ for the ray-tracing, on the emergent disk spectrum. The relative error is calculated as $\frac{F_{R_{\rm out}}-F_{2000R_g}}{F_{2000R_g}}$. $F_r$ denotes the X--ray flux calculated with $R_{\rm out}=r$. We use $F_{2000R_g}$ as the reference spectrum, as regions beyond $2000~R_g$ do not produce significant X--ray emissions due to the disk temperature decreasing as a function of radius. In the last test (bottom--middle), values of the disk parameters are similar to our best-fit results for AT2018cow (Table~\ref{tb:jointfit}).}
    \label{fig:testout}
\end{figure}

For a 10~$M_{\odot}$ BH disrupting a solar--type star, the TDE disk radius could be well above $1\times10^5$~$R_g$. However, the regions $\geq10^3$~$R_g$ contribute little to the X--ray flux, since the disk temperature decreases with radius. It is orders--of--magnitude lower for $\geq10^3$~$R_g$ compared to the innermost disk region \citep[e.g.,][]{straub2011testing,skadowski2011relativistic}. As it is computationally too expensive to ray--trace the whole disk, when constructing the spectral library for \texttt{slimdz}, we set a fixed value of the outer disk radius $R_{\rm out}$, and do not ray--trace the disk region further than $R_{\rm out}$. \citealt{wen2021mass} estimate a flux error of $\lesssim1\%$ when choosing $R_{\rm out}=600~R_g$ for a $10^4$~$M_{\odot}$ BH. Here we test different choices of $R_{\rm out}$ when ray--tracing the slim disk around a $10~M_{\odot}$ BH.

We consider a $10~M_{\odot}$, non--spinning BH with $\dot m=\dot m_{\rm Edd}$, observed at an inclination of $45^{\circ}$. Fig.~\ref{fig:nc} shows the relative flux differences between different $R_{\rm out}$ choices. The relative flux error is $\lesssim1\%$ when $R_{\rm out}>800~R_g$. For higher $M_{\bullet}$, higher $a_{\bullet}$, and lower $\dot m$, the flux difference would be smaller, as the innermost disk region becomes more dominant compared to the outer disk region. Therefore, in the main text, we produce the spectral library of \texttt{slimdz} with $R_{\rm out}=800~R_g$, accelerating the calculation while the flux error is minimal.

Note that a choice of $R_{\rm out}=800~R_g$ could introduce larger flux errors for a lower $a_{\bullet}$ or for a higher $\dot m$. In extreme cases, e.g., $M_{\bullet}=10~M_{\odot},~a_{\bullet}=-0.998,~\dot m=100~\dot m_{\rm Edd}$, the flux error could be as large as $\sim50\%$ (Fig.~\ref{fig:ec}). Therefore, one should be cautious when modeling a retrograde stellar--mass BH and a disk of large $\dot m$ with the currently available \texttt{slimdz} model.

Lastly, we estimate the flux error imposed by the choice of $R_{\rm out}=800~R_g$, in the particular case of AT2018cow. We consider the case of $M_{\bullet}=250~M_{\odot}$, $a_{\bullet}=0.4$, $\dot m=5~\dot m_{\rm Edd}$, and $\theta=74^{\circ}$ (as derived from the \texttt{slimdz} disk modeling in Table~\ref{tb:jointfit}). We find a relative flux error of $\lesssim0.5\%$ with $R_{\rm out}=800~R_g$ (Fig.~\ref{fig:oc}).

\section{Tables}

\begin{table*}
\centering
\caption{Journal listing properties of the observations analyzed in this paper.}
\scriptsize
\begin{tabular}{ccccccc}
\hline
Epoch & Satellite & ObsID(Label) & Date & Exposure (ks) & ${\rm R_{circ}}$ (\arcsec)& Energy band (keV) \\
\hline
Epoch 1 & \nus{} & 90401327002 & 2018-06-23 & 32.4 & 30 & 3.0-50.0~keV \\
 & \swf{} & 00010724012 & 2018-06-23 & 2.5 & 59 & 0.3-7.0 \\
 &  & 00010724013 & 2018-06-23 &  &  &  \\
 &  & 00010724010 & 2018-06-23 &  &  &  \\
\hline
Epoch 2 & \nus{} & 90401327004 & 2018-07-02 & 30.0 & 30 & 3.0-50.0 \\
 & \swf{} & 00010724046 & 2018-07-02 & 2.5 & 59 & 0.3-7.0 \\
 &  & 00010724047 & 2018-07-02 &  &  &  \\
 &  & 00010724048 & 2018-07-02 &  &  &  \\
 &  & 00010724049 & 2018-07-02 &  &  &  \\
\hline
Epoch 3 & \nus{} & 90401327006 & 2018-07-14 & 31.2 & 30 & 3.0-30.0 \\
 & \swf{} & 00088782001 & 2018-07-14 & 2.2 & 35.4 & 0.3-7.0 \\
\hline
Epoch 4 & \nus{} & 90401327008 & 2018-07-22 & 32.9 & 30\arcsec{} & 3.0-30.0 \\
 & \swf{} & 00010724087 & 2018-07-21 & 1.6 & 35.4 & 0.3-7.0 \\
 & \xmm{} & 0822580401 & 2018-07-23 & 33.0 & 25 & 0.3-10.0 \\
\hline
Epoch 5 & \xmm{} & 0822580501 & 2018-09-06 & 45.0 & 25 & 0.3-5.0 \\
\hline
Epoch 6 & \xmm{} & 0822580601 & 2019-01-20 & 56.4 & 25 & 0.3-1.5 \\
\hline
\end{tabular}
\tablefoot{We group all observations by time into 6 epochs and fit the data in each epoch simultaneously. \swf{} observations within the same epoch are combined and treated as a single observation. We also give the radius of the circular region that we use for the source$+$background spectral extraction in each observation (${\rm R_{circ}}$). We followed \citet{evans2009methods} in determining the radius of this source$+$background region to extract for \swf{} observations. In our spectral analysis, the energy band we use is given in the last column.}
\label{tb:obslist}
\end{table*}

\begin{table*}
\renewcommand{\arraystretch}{1.5}
\centering
\caption{Parameter constraints from fitting the data within each epoch with a power--law model.}
\scriptsize
\begin{tabular}{cc|cccccc}
    \hline
    Model & Parameter & Epoch 1 & Epoch 2 & Epoch 3 & Epoch 4 & Epoch 5 & Epoch 6 \\
    \hline
    \texttt{Constant} & $C_{\rm FPMA}$ & \multicolumn{6}{c}{[1.0
    ]}\\
    & $C_{\rm FPMB}$ & -
    & $1.01\pm0.03$ & $0.92\pm0.04$ & $1.00\pm0.06$ & - & -\\
    & $C_{\rm Swift}$ & -
    & $0.95\pm0.07$ & $0.47\pm0.08$ & $0.9\pm0.1$ & - & -\\
    & $C_{\rm XMM-Newton}$ & - & - & - & $0.78\pm0.04$ & - & -\\
    \hline
    \texttt{TBabs} & $N_{\rm H}$ ($10^{20}$~cm$^{-2}$) & \multicolumn{6}{c}{[5.0
    ]} \\
    \hline
    \texttt{powerlaw} & $\Gamma$ & -
    & $1.38\pm0.02$ & $1.52\pm0.05$ & $1.68\pm0.01$ & $1.7\pm0.2$ & $2.4\pm0.7$\\
    & norm (keV$^{-1}$cm$^{-2}$s$^{-1}$) & -
    & $(10.0\pm0.5)\times10^{-4}$ & $(4.2\pm0.4)\times10^{-4}$ & $(3.9\pm0.2)\times10^{-4}$ & $(3.7\pm0.4)\times10^{-6}$ & $(1.2\pm0.4)\times10^{-6}$\\
    \hline
    \multicolumn{2}{c|}{C-stat/d.o.f.} & -
    & 256.8/267 & 174.0/158 & 221.0/231 & 47.0/33 & 12.0/9 \\
    \hline
\end{tabular}
\tablefoot{We use a constant model to account for differences in normalization between each instrument and fix the constant for \nus{}/PFMA spectra $C_{\rm FPMA}=1$ in each epoch. The total fit function is thus \texttt{"constant*TBabs*powerlaw"}. Parameter values held fixed during the fit are given inside square brackets. No parameter constraints are given for Epoch 1, because at this epoch the source X--rays are dominated by reflection. With time the source spectrum seems to become softer.}
\label{tb:po}
\end{table*}

\begin{table*}
\renewcommand{\arraystretch}{1.5}
\centering
\caption{Parameter constraints from our spectral analysis, with a fit function of \texttt{"constant*TBabs*relxillCp"} for Epoch 1, and a fit function of \texttt{"constant*TBabs*(powerlaw+slimdz)"} for Epoch 2 to 6.}
\scriptsize
\begin{tabular}{cc|cccccc}
    \hline
    Model & Parameter & Epoch 1 & Epoch 2 & Epoch 3 & Epoch 4 & Epoch 5 & Epoch 6 \\
    \hline
    \texttt{Constant} & $C_{\rm FPMB}$ & $0.96\pm0.02$ & $1.01\pm0.03$ & $0.92\pm0.04$ & $1.00\pm0.06$ & - & -\\
    & $C_{\rm Swift}$ & $1.00\pm0.06$ & $0.95\pm0.07$ & $0.47\pm0.08$ & $0.9\pm0.1$ & - & -\\
    & $C_{\rm XMM-Newton}$ & - & - & - & $0.79\pm0.04$ & - & -\\
    \hline
    \texttt{TBabs} & $N_{\rm H}$ ($10^{20}$~cm$^{-2}$) & \multicolumn{6}{c}{[5.0
    ]} \\
    \hline
    \texttt{powerlaw} & $\Gamma$ & - & $1.38\pm0.02$ & $1.52\pm0.05$ & $1.52\pm0.05$ & $1.7\pm0.2$ & -\\
    & norm (keV$^{-1}$cm$^{-2}$s$^{-1}$) & - & $(10.0\pm0.5)\times10^{-4}$ & $(4.2\pm0.4)\times10^{-4}$ & $(2.8\pm0.4)\times10^{-4}$ & $(3.7\pm0.4)\times10^{-6}$ & -\\
    \hline
    \texttt{slimdz} & $\dot m$ ($\dot m_{\rm Edd}$) & - & - & - & $>$0.5 & - & $<$4.0\\
    & $\theta$ ($^{\circ}$)& - & - & - & $<$81 & - & NC\\
    & log($M_{\bullet}$/$M_{\odot}$) & - & - & - & $2.4_{-0.2}^{+0.8}$ & - & $2.3\pm0.9$\\
    & $a_{\bullet}$ & - & - & - & $>$-0.6 & - & NC\\
    \hline
    \texttt{relxillCp} & $\theta$ ($^{\circ}$)& $74\pm2$ & - & - & - & - & -\\
    & $a_{\bullet}$ & $0.98\pm0.01$ & - & - & - & - & -\\
    & $kT_e$ (keV) & $28\pm8$ & - & - & - & - & -\\
    & $\Gamma$ & $1.22\pm0.02$ & - & - & - & - & -\\
    & $q$ & $4\pm2$ & - & - & - & - & -\\
    & log($\xi$) & $3.45\pm0.06$ & - & - & - & - & -\\
    & $A_{\rm Fe}$ & $>$8.0 & - & - & - & - & -\\
    & Refl$_{\rm frac}$ & $11_{-4}^{+21}$ & - & - & - & - & -\\
    & log($\rho$/cm$^{-3}$) & $<$15.2 & - & - & - & - & -\\
    & norm (erg~cm$^{-2}$s$^{-1}$) & $(3\pm2)\times10^{-5}$ & - & - & - & - & -\\
    \hline
    \multicolumn{2}{c|}{C-stat/d.o.f.} & 279.5/275 & 256.8/267 & 174.0/158 & 200.2/227 & 47.0/33 & 12.2/7 \\
    \hline
\end{tabular}
\tablefoot{Parameter values held fixed during the fit
are given inside square brackets. We use a constant model to account for differences in normalization between each instrument and fix the constant for \nus{}/PFMA spectra $C_{\rm FPMA}=1$ in each epoch. For the models \texttt{slimdz} and \texttt{relxillCp}, the redshift of AT2018cow ($z=0.01404$) is taken. We find that only for Epoch 4 adding a slim disk component significantly improves the fit compared to only using a power--law model (Table~\ref{tb:po}). Nonetheless, we also test a model of a slim disk alone (\texttt{"constant*TBabs*slimdz"}) for Epoch 6, as the source becomes much softer at this epoch compared to previous epochs. The symbol "NC" means the parameter cannot be constrained within the allowed range of values (for $\theta$ this is $3^{\circ}\leq\theta\leq90^{\circ}$, and for $a_{\bullet}$ this is $-0.998\leq~a_{\bullet}\leq0.998$). See \citealt{dauser2014role} and \citealt{garcia2014improved} for detailed descriptions of \texttt{relxillCp} parameters.}
\label{tb:1to6}
\end{table*}

\begin{table*}
\renewcommand{\arraystretch}{1.5}
\centering
\caption{Same as Table~\ref{tb:1to6}, but here we jointly fit all data from Epoch 2 to 6.}
\scriptsize
\begin{tabular}{cc|cccccc}
    \hline
    Model & Parameter & Epoch 1 & Epoch 2 & Epoch 3 & Epoch 4 & Epoch 5 & Epoch 6 \\
    \hline
    \texttt{Constant} & $C_{\rm FPMB}$ & $0.96\pm0.02$ & $1.01\pm0.03$ & $0.92\pm0.04$ & $1.00\pm0.06$ & - & -\\
    & $C_{\rm Swift}$ & $1.00\pm0.05$ & $0.95\pm0.07$ & $0.46\pm0.08$ & $0.9\pm0.1$ & - & -\\
    & $C_{\rm XMM-Newton}$ & - & - & - & $0.79\pm0.04$ & - & -\\
    \hline
    \texttt{TBabs} & $N_{\rm H}$ ($10^{20}$~cm$^{-2}$) & \multicolumn{6}{c}{[5.0
    ]} \\
    \hline
    \texttt{powerlaw} & $\Gamma$ & - & $1.38\pm0.02$ & $1.52\pm0.05$ & $1.51\pm0.05$ & $0.9_{-1.1}^{+0.6}$ & -\\
    & norm (keV$^{-1}$cm$^{-2}$s$^{-1}$) & - & $(10.0\pm0.5)\times10^{-4}$ & $(4.2\pm0.4)\times10^{-4}$ & $(2.7\pm0.3)\times10^{-4}$ & $(2.1\pm1.5)\times10^{-6}$ & -\\
    \hline
    \texttt{slimdz} & $\dot m$ ($\dot m_{\rm Edd}$) & - & $<$4.9 & $<$18 & $>$4.8 & $<$0.23 & $<$0.17 \\
    & $\theta$ ($^{\circ}$)& - & \multicolumn{5}{c}{$<$76}\\
    & log($M_{\bullet}$/$M_{\odot}$) & - & \multicolumn{5}{c}{$2.4_{-0.1}^{+0.6}$}\\
    & $a_{\bullet}$ & - & \multicolumn{5}{c}{$0.4_{-0.9}^{+0.5}$}\\
    \hline
    \texttt{relxillCp} & $\theta$ ($^{\circ}$)& $74\pm2$ & - & - & - & - & -\\
    & $a_{\bullet}$ & $0.98\pm0.01$ & - & - & - & - & -\\
    & $kT_e$ (keV) & $28\pm8$ & - & - & - & - & -\\
    & $\Gamma$ & $1.22\pm0.02$ & - & - & - & - & -\\
    & $q$ & $4\pm2$ & - & - & - & - & -\\
    & log($\xi$) & $3.45\pm0.06$ & - & - & - & - & -\\
    & $A_{\rm Fe}$ & $>$8.0 & - & - & - & - & -\\
    & Refl$_{\rm frac}$ & $11_{-4}^{+21}$ & - & - & - & - & -\\
    & log($\rho$/cm$^{-3}$) & $<$15.2 & - & - & - & - & -\\
    & norm (erg~cm$^{-2}$s$^{-1}$) & $(3\pm2)\times10^{-5}$ & - & - & - & - & -\\
    \hline
    \multicolumn{2}{c|}{C-stat/d.o.f.} & 279.5/275 & \multicolumn{5}{c}{688.5/691} \\
    \hline
\end{tabular}
\tablefoot{We force the slim disk component in each epoch to have the same inclination $\theta$, BH mass $M_{\bullet}$, and BH spin $a_{\bullet}$. Since statistically the slim disk component is not needed for Epoch 2, 3, and 5, only an upper limit on the disk accretion rate was obtained for those epochs.}
\label{tb:jointfit}
\end{table*}

\begin{table*}
\renewcommand{\arraystretch}{1.5}
\centering
\caption{Same as Table~\ref{tb:1to6}, but here we jointly fit all data from Epoch 2 to 6, and we replace \texttt{slimdz} with \texttt{slimbh}.}
\scriptsize
\begin{tabular}{cc|cccccc}
    \hline
    Model & Parameter & Epoch 1 & Epoch 2 & Epoch 3 & Epoch 4 & Epoch 5 & Epoch 6 \\
    \hline
    \texttt{Constant} & $C_{\rm FPMB}$ & $0.96\pm0.02$ & $1.01\pm0.03$ & $0.92\pm0.05$ & $1.00\pm0.06$ & - & -\\
    & $C_{\rm Swift}$ & $1.00\pm0.05$ & $0.94\pm0.07$ & $0.47\pm0.09$ & $0.9\pm0.1$ & - & -\\
    & $C_{\rm XMM-Newton}$ & - & - & - & $0.79\pm0.04$ & - & -\\
    \hline
    \texttt{TBabs} & $N_{\rm H}$ ($10^{20}$~cm$^{-2}$) & \multicolumn{6}{c}{[5.0
    ]} \\
    \hline
    \texttt{powerlaw} & $\Gamma$ & - & $1.38\pm0.02$ & $1.52\pm0.05$ & $1.57\pm0.04$ & $1.1\pm0.3$ & -\\
    & norm (keV$^{-1}$cm$^{-2}$s$^{-1}$) & - & $(10.0\pm0.5)\times10^{-4}$ & $(4.2\pm0.4)\times10^{-4}$ & $(3.1\pm0.2)\times10^{-4}$ & $(2.5^{+0.4}_{-0.7})\times10^{-6}$ & -\\
    \hline
    \texttt{slimbh} & $L_{\rm disk}$ ($L_{\rm Edd}$) & - & $<$1.2 & NC & $>$1.0 & $<$0.07 & $<$0.06 \\
    & $\theta$ ($^{\circ}$)& - & \multicolumn{5}{c}{$75\pm8$}\\
    & log($M_{\bullet}$/$M_{\odot}$) & - & \multicolumn{5}{c}{$3.2\pm0.2$}\\
    & $a_{\bullet}$ & - & \multicolumn{5}{c}{$<0.5$}\\
    \hline
    \texttt{relxillCp} & $\theta$ ($^{\circ}$)& $74\pm2$ & - & - & - & - & -\\
    & $a_{\bullet}$ & $0.98\pm0.01$ & - & - & - & - & -\\
    & $kT_e$ (keV) & $28\pm8$ & - & - & - & - & -\\
    & $\Gamma$ & $1.22\pm0.02$ & - & - & - & - & -\\
    & $q$ & $4\pm2$ & - & - & - & - & -\\
    & log($\xi$) & $3.45\pm0.06$ & - & - & - & - & -\\
    & $A_{\rm Fe}$ & $>$8.0 & - & - & - & - & -\\
    & Refl$_{\rm frac}$ & $11_{-4}^{+21}$ & - & - & - & - & -\\
    & log($\rho$/cm$^{-3}$) & $<$15.2 & - & - & - & - & -\\
    & norm (erg~cm$^{-2}$s$^{-1}$) & $(3\pm2)\times10^{-5}$ & - & - & - & - & -\\
    \hline
    \multicolumn{2}{c|}{C-stat/d.o.f.} & 279.5/275 & \multicolumn{5}{c}{691.8/691} \\
    \hline
\end{tabular}
\tablefoot{We force the slim disk component in each epoch to have the same inclination $\theta$, BH mass $M_{\bullet}$, and BH spin $a_{\bullet}$. Furthermore, we replace \texttt{slimdz} with a pre--existing slim disk model \texttt{slimbh}. The total fit function for the joint--fit is then \texttt{"constant*Tbabse*(powerlaw+slimbh)"}. See the main text for the physical differences between the models. Note in \texttt{slimbh}, $0.05<L_{\rm disk}/L_{\rm Edd}<1.25$ with $L_{\rm Edd}\equiv1.26\times10^{38}(M/M_{\odot})$~erg/s, and only the prograde spin is considered ($a_{\bullet}>0$). Both limb--darkening and surface profile are switched--on in \texttt{slimbh} during the fit (model switch vflag$=1$ and lflag$=1$), and the disk viscosity parameter $\alpha=0.1$.}
\label{tb:slimbh}
\end{table*}

\begin{table*}
\renewcommand{\arraystretch}{1.5}
\centering
\caption{Joint--fits of \xmm{}/EPIC-pn spectra from Epoch 5 and 6.}
\scriptsize
\begin{tabular}{cc|cc||cc|cc}
    \hline
    Model & Parameter & Epoch 5 & Epoch 6 & Model & Parameter & Epoch 5 & Epoch 6\\
    \hline
    \texttt{TBabs} & $N_{\rm H}$ ($10^{20}$~cm$^{-2}$) & \multicolumn{2}{c||}{[5.0
    ]} & \texttt{TBabs} & $N_{\rm H}$ ($10^{20}$~cm$^{-2}$) & \multicolumn{2}{c}{[5.0
    ]} \\
    \hline
    \texttt{powerlaw} & $\Gamma$ & $0.9_{-3.1}^{+0.8}$ & - & \texttt{powerlaw} & $\Gamma$ & $0.7_{-3.2}^{+0.8}$ & - \\
    & norm (keV$^{-1}$cm$^{-2}$s$^{-1}$) & $(190_{-188}^{+171})\times10^{-8}$ & - & & norm (keV$^{-1}$cm$^{-2}$s$^{-1}$) & $(151_{-149}^{+181})\times10^{-8}$ & -\\
    \hline
    \texttt{slimdz} & $\dot m$ ($\dot m_{\rm Edd}$) & $<$4.6 & $<$3.6 & \texttt{slimbh} & $L_{\rm disk}$ ($L_{\rm Edd}$) & $<$0.60 & $<$0.17\\
    & $\theta$ ($^{\circ}$)& \multicolumn{2}{c||}{NC \{$3^{\circ}\leq\theta\leq90^{\circ}$\}} & & $\theta$ ($^{\circ}$)& \multicolumn{2}{c}{NC \{$0^{\circ}\leq\theta\leq85^{\circ}$\}} \\
    & log($M_{\bullet}$/$M_{\odot}$) & \multicolumn{2}{c||}{$2.5_{-0.9}^{+1.4}$} & & log($M_{\bullet}$/$M_{\odot}$) & \multicolumn{2}{c}{$2.2_{-0.2}^{+1.0}$} \\
    & $a_{\bullet}$ & \multicolumn{2}{c||}{NC \{$-0.998\leq~a_{\bullet}\leq0.998$\}}& & $a_{\bullet}$ & \multicolumn{2}{c}{NC \{$0<~a_{\bullet}<0.999$\}}\\
    \hline
    \multicolumn{2}{c|}{C-stat/d.o.f.} & \multicolumn{2}{c||}{57.7/39} & \multicolumn{2}{c|}{C-stat/d.o.f.} & \multicolumn{2}{c}{57.3/39}\\
    \hline
\end{tabular}
\tablefoot{At these two epochs, the disk is likely to be at sub--Eddington mass accretion rates. We compare the results using the \texttt{slimdz} model with those using the \texttt{slimbh} model, employing a fit function of \texttt{"constant*TBabs*(powerlaw+slimdz)"} and a fit function of \texttt{"constant*TBabs*(powerlaw+slimbh)"}, respectively. The symbol "NC" means the parameter cannot be constrained within the range of values allowed by models (listed in curly brackets).}
\label{tb:last2}
\end{table*}

\newpage
\label{lastpage}
\end{document}